\documentclass[final,5p,times,twocolumn]{elsarticle}

\usepackage{graphics}
\usepackage{graphicx}

\usepackage{amssymb}

\journal{Physica C}

\begin{document}
\newcommand*{\cm}{cm$^{-1}$\,}
\newcommand*{\Tc}{T$_c$\,}

\begin{frontmatter}



\title{Optical and Raman spectroscopy
studies on Fe-based superconductors}


\author[label1]{W. Z. Hu}
\author[label2,label3]{Q. M. Zhang}
\author[label1]{N. L. Wang\corref{cor1}}
\ead{nlwang@aphy.iphy.ac.cn Tel: +86-10-8264 9584.}

\address[label1]{Beijing National Laboratory for Condensed Matter
Physics, Institute of Physics, Chinese Academy of Sciences,
Beijing 100190, P. R. China}

\address[label2]{Department of Physics, Renmin University of China, Beijing 100872,
P. R. China}

\address[label3]{National Laboratory of Solid
State Microstructures, Department of Physics, Nanjing University,
Nanjing 210093, P. R. China}

\begin{abstract}
A brief review of optical and Raman studies on the Fe-based
superconductors is given, with special emphasis on the competing
phenomenon in this system. Optical investigations on ReFeAsO
(Re=rare-earth element) and AFe$_2$As$_2$ (A=alkaline-earth metal)
families provide clear evidence for the gap formation in the
broken symmetry states, including the partial gaps in the
spin-density wave states of parent compounds, and the pairing gaps
in the superconducting states for doped compounds. Especially, the
superconducting gap has an s-wave pairing lineshape in hole-doped
BaFe$_2$As$_2$. Optical phonons at zone center detected by Raman
and infrared techniques are classified for several Fe-based
compounds. Related issues, such as the electron-phonon coupling
and the effect of spin-density wave and superconducting
transitions on phonons, are also discussed. Meanwhile, open
questions including the \emph{T}-dependent mid-infrared peak at
0.6-0.7 eV, electronic correlation, and the
similarities/differences between high-\Tc cuprates and Fe-based
superconductors are also briefly discussed. Important results from
other experimental probes are compared with optical data to better
understand the spin-density wave properties, the
superconductivity, and the multi-band character in Fe-based
compounds.
\end{abstract}


\end{frontmatter}


\section{Introduction}
\label{intro}

A new family of high temperature superconductors containing FeAs
layers has attracted a broad interest since the discovery of
superconductivity in LaFeAsO$_{1-x}$F$_x$ (x=0.05$\sim$0.12) with
$T_c$=26 K by Kamihara \emph{et al}.\cite{Kamihara08} When
replacing the La ion by other rare-earth elements (Ce, Nd, Pr,
Sm), $T_c$ could be raised to 40$\sim$55 K.\cite{XHChen,
ChenPRLCe, Ren1} In analogy with the high-\Tc cuprates, those
quaternary oxypnictides adopt a layered structure with the
Fe$_2$As$_2$ tetrahedron as an essential structural unit, in which
the superconductivity appears when the antiferromagnetic ordering
for this Fe-layer is suppressed by doping. However, unlike the
cuprates where the undoped parent compounds are antiferromagnetic
Mott insulators, the ReFeAsO (Re=La, Ce, Nd, Pr, Sm, etc.) parent
compounds are metals or semimetals. The magnetic phase for the
Fe-based parent compounds is identified as a spin-density wave
(SDW) ordering.\cite{DongEPL,Clarina}

Although the first FeAs superconductors are found on doped
variants of ReFeAsO, it is soon reported that superconductivity
could be induced in other types of structure containing the same
tetrahedrally coordinated Fe$_2$As$_2$ (or Fe$_2$Se$_2$) layers as
in ReFeAsO compounds. Those include the oxygen-free AFe$_2$As$_2$
(A=Ca, Sr, Ba, Eu) (122-type)
\cite{Rotter1,Rotter2,Krellner,Chen2,Huang,Zhao,Ni,Chen3,Jeevan,ZhiRen},
FeSe(Te) (11-type) \cite{FeSe,FeSeTe} and Li$_x$FeAs (111-type)
\cite{CQJin,Tapp,Pitcher}. Up to now, many of research works have
been done on 122-type materials. This is because high-quality
single crystals with sufficient sizes for different measurements
could be easily grown for this type of Fe-based compounds,
although their maximum \Tc does not exceed 38 K.

Optical and Raman spectroscopies are powerful experimental
techniques in determining the charge dynamics and phonon response
for solids. In this article, we briefly summarize current progress
in optical and Raman spectroscopic studies, including the normal
state properties, the SDW partial gap and superconducting gap,
electronic correlation, and phonon spectra for Fe-based compounds.
Comprehensive overviews on various physical properties for the
superconducting Fe-based compounds could be found in this special
issue on pnictide superconductors or in a recent review article by
Sadovskii. \cite{Sadovskii}

In the early stage, infrared spectroscopic studies on
poly-crystalline Fe-based compounds provide primitive information
for the energy gaps in the broken symmetry states. The SDW partial
gap is evidenced as a spectral suppression in the far-infrared for
LaFeAsO, with a metallic response near the low frequency
limit.\cite{DongEPL,ChenPRLCe} Optical signature of the
superconducting gap is also observed in superconducting
ReFeAsO$_{1-x}$F$_x$ polycrystals\cite{ChenPRLLa, Dubroka}. When
high-quality single crystalline AFe$_2$As$_2$ becomes available,
convinced spectroscopic evidence are obtained for the in-plane
charge dynamics. Direct observation of the SDW partial gap in
AFe$_2$As$_2$, and the s-wave symmetry of superconducting gap in
Ba$_{0.6}$K$_{0.4}$Fe$_2$As$_2$ are reported.\cite{Hu122, BaKPRL}
Optical evidence for the s-wave superconducting gap indicates that
the newly discovered Fe-based superconductor is apparently
distinguishable from the high-\Tc cuprates. More interestingly,
both the SDW gap in parent compounds and the superconducting gap
in the doped superconductors show a double-gap character for
AFe$_2$As$_2$-type compound, reflecting the multi-band property in
Fe-based system. Spectroscopic studies on single crystalline 122-
and 1111-type Fe-based superconductors suggest the electronic
correlation is moderate, with a possible electron-boson coupling
signature in the optical quasiparticle self-energy, but the
scattering rate 1/$\tau(\omega)$ shows apparently different
response to that of high-\Tc cuprates. A 0.6-0.7 eV mid-infrared
feature is observed for both ReFeAsO and AFe$_2$As$_2$ systems,
with a spectral weight redistribution to higher energies as
decreasing \emph{T}. Whether this high energy feature is a
pseudogap or just an interband transition remains an open
question.

The resistivity and free-carrier concentration of iron-based
superconductors and their parent compounds are comparable to those
of cuprates, which can be estimated by the published transport,
Hall and infrared measurements and first-principle calculations.
However, both experiments and theoretical calculations indicate
that the Fe-based compounds show multi-band structures of
\emph{d}-orbitals, which can effectively increase the itinerancy
of electrons and decrease electronic correlation. Consequently
this would lead to a relatively enhanced Coulomb screening, which
may raise the difficulties in searching for electronic Raman
scattering signals. As we know, electronic Raman scattering
detects density-density correlation in the charge channel. So far,
studies on the zone center optical phonons obtain consistent
results, while the ratio of signal to noise is low in Raman
measurements compared to the case of cuprates. Raman and infrared
active modes are investigated in combination with first-principle
calculations. Compatible results are obtained by different groups.
Raman, inelastic X-ray scattering measurements, and isotope effect
experiments etc. provide important clues on the electron-phonon
coupling in the Fe-based materials. Considering the similarities
between the antiferromagnetic state in cuprates and the SDW state
in the parent compounds of iron-based superconductors, it will be
very interesting to explore if some magnetic excitations similar
to two-magnon process can be observed by inelastic light
scattering. Unfortunately, no such magnetic Raman scattering is
reported at present.

\section{Spin-density wave in the parent compound} \label{SDWgap}

An important progress in understanding the FeAs-based compounds is
the identification of the SDW order in the parent compound. As we
shall explain, essential information was obtained from the
infrared spectroscopy measurement.

\subsection{SDW in ReFeAsO}\label{1111SDW}

\begin{figure}[t]
\begin{center}
\includegraphics[width=2.7in]{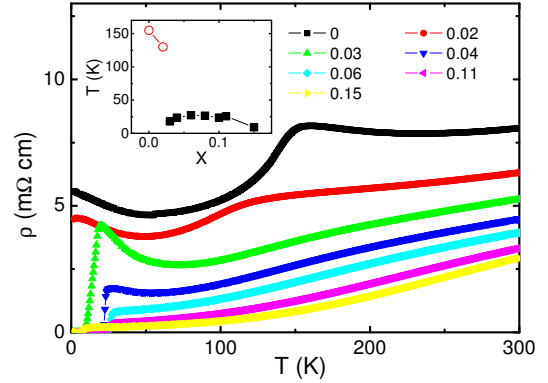}%
\vspace*{-0.20cm}%
\caption{\label{fig:Resistivity} The temperature dependent
resistivity for a series of LaFeAsO$_{1-x}$F$_x$ polycrystals.
Inset: The phase diagram showing the anomaly (red circle) and
superconducting transition (black square) temperatures as a
function of fluorine content. \cite{DongEPL}}
\end{center}
\end{figure}

The FeAs-based system shows an interesting competing phenomenon.
As shown in Fig. \ref{fig:Resistivity}, the undoped LaFeAsO
compound itself is not superconducting, but shows a strong anomaly
near 150 K, below which the resistivity drops steeply. With
fluorine doping, the anomaly shifts to lower temperature, and
gradually disappears, then superconducting transition occurs.
Apparently, superconductivity competes with the phase showing the
anomaly in this system. The anomaly was already observed in the
earliest report by Kamihara et al.\cite{Kamihara08}, however its
origin was not addressed there. The first work of identifying the
nature of this anomaly was done by Dong \emph{et al.} in a
combined experimental studies of transport/optical properties and
first-principle band structure calculations.\cite{DongEPL} Dong
\emph{et al.} found a very clear specific heat jump from specific
heat measurement, suggesting that the anomaly is a second-order
phase transition. Furthermore, the electronic specific heat
coefficient determined at low temperature, $\gamma$=3.7 mJ/mol
K$^2$, is significantly smaller than the value obtained from the
band structure calculations (about 5.5-6.5 mJ/mol K$^2$). Such a
discrepancy is unconventional, because the band structure
calculation usually gives a smaller value than the experimental
data, and their difference is ascribed to the renormalization
effect. Here a natural explanation for a larger theoretically
predicted value is that a partial energy gap opens, which removes
parts of the density of states below the phase transition
temperature. This is indeed seen in their optical spectroscopy
experiments on the parent compound.

\begin{figure}[t]
\begin{center}
\includegraphics[width=2.7in]{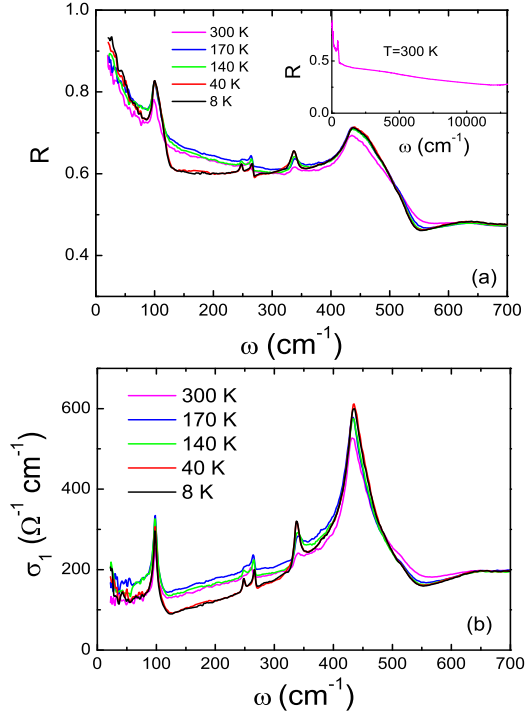}%
\vspace*{-0.20cm}%
\caption{\label{fig:LaFeAsOEPL} The optical reflectivity
R($\omega$) and the real part of conductivity for LaFeAsO
polycrystal.\cite{DongEPL}}
\end{center}
\end{figure}

Figure \ref{fig:LaFeAsOEPL} shows the far-infrared reflectivity
R($\omega$) and conductivity $\sigma_1(\omega)$ for
polycrystalline LaFeAsO.\cite{DongEPL} Regardless of the five
phonon modes in the far-infrared, one can find the reflectance is
strongly suppressed below 600 cm$^{-1}$ at low temperature. This
leads to a loss of low-frequency spectral weight in the real part
of conductivity $\sigma_1(\omega)$ as shown in Fig.
\ref{fig:LaFeAsOEPL}b. In fact, the $\sigma_1(\omega)$ already
shows a weak suppression at 140 K, which becomes more apparent as
decreasing temperature. A spectral suppression in the free-carrier
contribution indicates a reduction of itinerant carriers, thus is
an optical evidence for the energy gap on the Fermi surface. As
the reflectance at very low frequency increases fast and exceeds
the values at high temperature, the compound is still metallic
even below the phase transition, being consistent with the
enhanced \emph{dc} conductivity. The data indicate clearly that
the Fermi surfaces are only partially gapped.

Identification of a gap formation below the phase transition
strongly suggests that the anomaly is caused by a
symmetry-breaking phase transition, i.e. either a charge-density
wave (CDW) or an SDW instability. In the far-infrared spectra, one
can see several infrared-active phonon modes roughly at 100, 247,
265, 336, 438 \cm, but no splitting or new phonon modes develop
below the transition. This indicates that the structural
distortion associated with the transition, if any, should be
rather small. On this basis, Dong \emph{et al.} suggest that the
phase transition is unlikely to be driven by a CDW instability but
an SDW instability. Further convincing evidence for an SDW
instability comes from the first-principle band structure
calculations presented in the same work,\cite{DongEPL} where the
authors find a strong nesting effect between the hole Fermi
surface surrounding the zone center and the electron Fermi surface
centered at the zone corner.

The Fermi surface nesting in LaFeAsO has a commensurate wave
vector \textbf{q}=($\pi ,\pi$), leading to a doubling of the unit
cell of $\sqrt{2}\times\sqrt{2}$ along diagonal directions.
Suppose that an antiferromagnetic ordering is associated with the
lattice structure, a stripe- or collinear-type antiferromagnetic
pattern would develop below the phase transition temperature with
such a ($\pi ,\pi$) nesting vector.\cite{DongEPL} Further
calculation indicates that the LaFeAsO parent compound would have
a lower total energy in such antiferromagnetic ordered phase with
respect to the nonmagnetic phase. Furthermore, the density of
state at the Fermi level is reduced significantly compared with
the original nonmagnetic phase, in accordance with a small
specific heat value from transport measurement, and a gap from
optical spectroscopy. Therefore, the SDW order for the parent
compound is identified convincingly.

Conclusive evidence for the spin-density wave order is obtained
from neutron experiments. Clarina de la Cruz \emph{et
al.}\cite{Clarina} performed neutron diffraction measurement on
LaFeAsO and demonstrated that an antiferromagnetic ordering
develops at low temperature. In particular, the spin structure
determined by their neutron experiment is identical to the spin
configuration predicted based on the nesting of hole and electron
Fermi surfaces proposed by Dong \emph{et al.}\cite{DongEPL}
However, the neutron experiments also revealed a very subtle
structural distortion near 150 K, which accounts for the
resistivity anomaly, while a magnetic ordering develops at a
slightly lower temperature for LaFeAsO polycrystal. It is now
widely believed that the structural distortion is driven by the
magnetic instability.\cite{Yildirim, Ma, Fang, Xu}

Optical reflectance investigations were also carried out on other
1111-type polycrystalline parent
compounds.\cite{ChenPRLCe,Marini,Tropeano} Shortly after the work
on LaFeAsO$_{1-x}$F$_x$, Chen \emph{et al.}\cite{ChenPRLCe} used
the rare-earth element Ce to replace La and synthesize a series of
CeFeAsO$_{1-x}$F$_x$ compounds. Interestingly, in this rare-earth
based 1111 compound, a much higher \Tc (41 K) is achieved. From
transport and optical spectroscopy measurement they identified
very similar competing phenomenon between spin-density wave and
superconductivity in CeFeAsO$_{1-x}$F$_x$. Neutron experiment also
revealed a stripe type antiferromagnetic order of the Fe
sublattice in the parent compound, being identical to the LaFeAsO
compound.\cite{ZhaoNatMat} The similarity between La- and Ce-based
compounds suggest that the interplay between the superconductivity
and the spin-density wave instability is a common phenomenon for
FeAs-based systems. Additionally Chen \emph{et al.} found that the
Ce 4\emph{f} electrons form local moments and ordered
antiferromagnetically below 4 K, which coexists with
superconductivity in CeFeAsO$_{1-x}$F$_x$.

\begin{figure}[t]
\begin{center}
\includegraphics[width=2.5in]{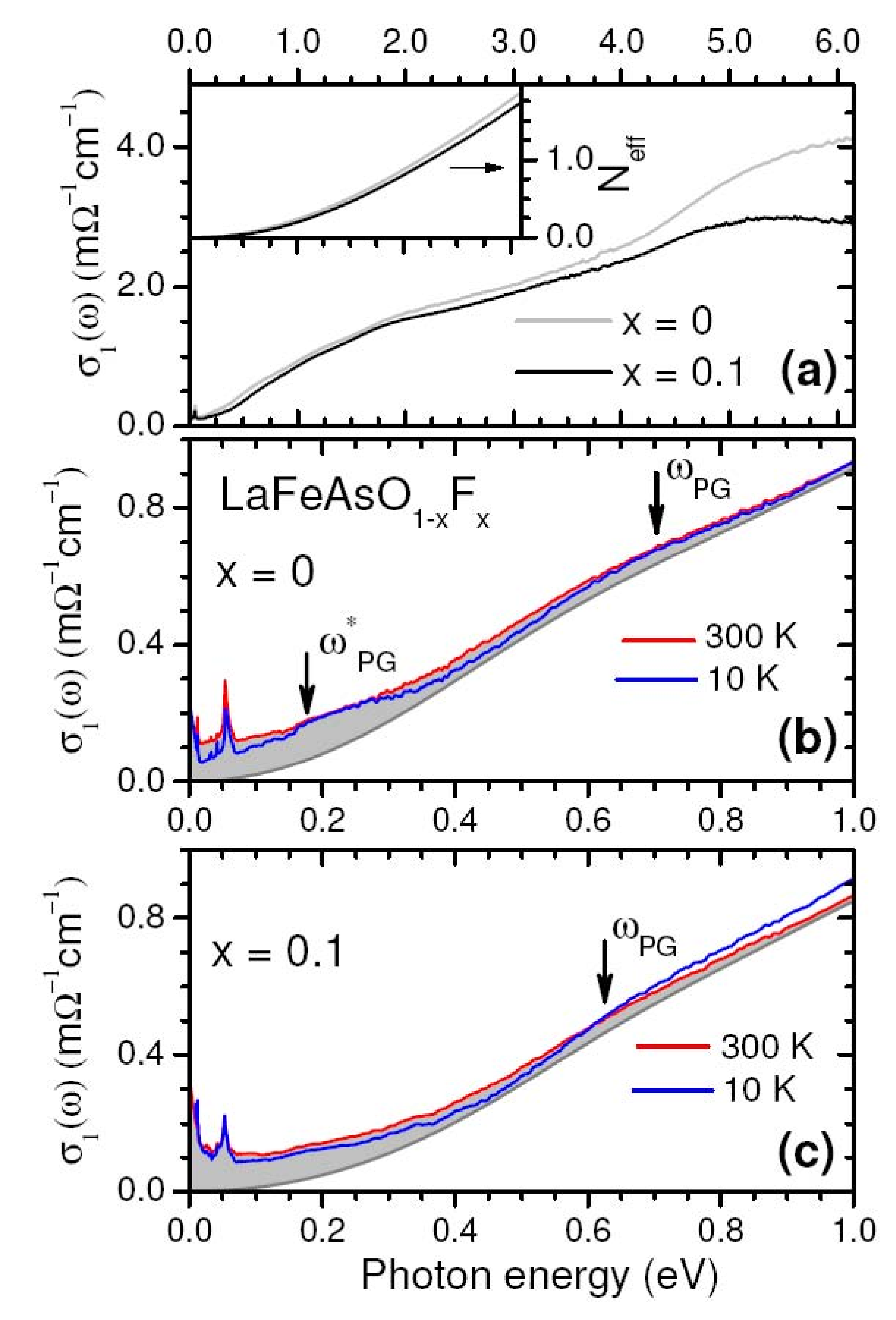}%
\vspace*{-0.20cm}%
\caption{\label{fig:BorisPG} (a) The optical conductivity
$\sigma_1(\omega)$ for LaFeAsO (gray) and LaFeAsO$_{0.9}$F$_{0.1}$
(black) at T=300 K. The $\sigma_1(\omega)$ below 1 eV for (b)
LaFeAsO$_{0.9}$F$_{0.1}$ and (c) LaFeAsO at 10 K and 300 K. The
dark gray lines is the low-energy tail of the interband
transitions. The shaded area is the contribution from intraband
response. The pseudogap is marked by arrows.\cite{BorisLa}}
\end{center}
\end{figure}

The optical suppression (i.e., the SDW gap) is found for different
types of ReFeAsO samples, however, as the optical data are
collected on polycrystals, the R($\omega$) is dominated by several
strong phonon modes related to the weakly conducting c-axis
response.\cite{Marini} In particular, a strong oxygen-derived
phonon mode\cite{Singh237003,Boeri} around 438 \cm weakens the
ab-plane charge dynamics, it is therefore difficult to accurately
define the energy scale affected by the SDW gap. Boris \emph{et
al.}\cite{BorisLa} report an infrared ellipsometry study on
polycrystalline LaFeAsO parent compound and 10\% F-doped
superconductor LaFeAsO$_{1-x}$F$_x$, and found that the
SDW-induced suppression in the optical conductivity
$\sigma_1(\omega)$ has a rather large energy scale.  Figure
\ref{fig:BorisPG} shows the optical conductivity for both the
parent and the F-doped samples. There is a low energy feature
around 0.16 eV in LaFeAsO ($\omega_{PG}^*$ in Fig.
\ref{fig:BorisPG}b), and the $\sigma_1(\omega)$ for 10 K is
suppressed below that of 300 K for $\omega<\omega_{PG}^*$. This
0.16 eV feature disappears in the superconducting compound (Fig.
\ref{fig:BorisPG}c). As F-doping suppress the antiferromagnetic
transition, the 0.16 eV feature in LaFeAsO should be the SDW gap
for the parent compound. Our published data on
LaFeAsO\cite{DongEPL} only show the spectra below 700 \cm (see
Fig. \ref{fig:LaFeAsOEPL}), it is not clear whether the energy
scope for the SDW-gap induced suppression extends to higher
energies. In our latter study on another LaFeAsO polycrystal with
higher quality (Fig. \ref{fig:EuFe2As2}), the optical suppression
indeed extend towards high frequency region, in agreement with the
energy scale of 0.16 eV observed by Boris \emph{et al}.

Boris \emph{et al.} report another gap-like feature around 0.65 eV
for both LaFeAsO and F-doped samples (see Fig. \ref{fig:BorisPG}b
and c), which is suggested to be a pseudogap.\cite{BorisLa}
Meanwhile, there is a significant spectral weight transfer for the
parent and F-doped compounds. In LaFeAsO, the spectral weight
redistribution associated with the pseudogap at 0.65 eV is
restricted below 2 eV; while for superconducting
LaFeAsO$_{0.9}$F$_{0.1}$, a substantial spectral weight transfer
from below 2 eV to above 4 eV with increasing temperature is
observed.\cite{BorisLa}

\subsection{SDW in AFe$_2$As$_2$}\label{122SDW}
\begin{figure}[b]
\begin{center}
\includegraphics[width=2.5in]{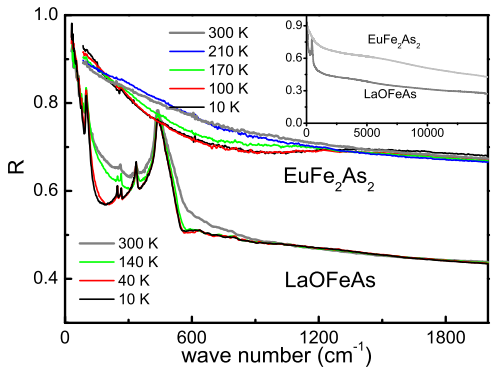}%
\vspace*{-0.20cm}%
\caption{\label{fig:EuFe2As2} The optical reflectivity R($\omega$)
for EuFe$_2$As$_2$ and LaFeAsO polycrystals.\cite{HuUnpub}}
\end{center}
\end{figure}

After numerous attempts in Re-site chemical substitution for
ReFeAsO, the maximum \Tc for this 1111-family is still below 56 K.
Therefore, finding new structures containing the FeAs layer
becomes the central task in materials research.
ThCr$_2$Si$_2$-type compound AFe$_2$As$_2$ (A=Ba, Sr, Eu) is one
of the important findings in the exploration for new types of
Fe-based compound.\cite{Rotter1} Figure \ref{fig:EuFe2As2}
compares the optical reflectivity for polycrystalline LaFeAsO and
EuFe$_2$As$_2$ (T$_{SDW}$=190 K).\cite{HuUnpub} One can find the
spectral suppression related to SDW partial gap has a larger
energy scale for EuFe$_2$As$_2$ than LaFeAsO. This could be due to
the higher SDW transition temperature for EuFe$_2$As$_2$. In
addition, the overall R($\omega$) for EuFe$_2$As$_2$ is
substantially higher than that of LaFeAsO, suggesting a larger
optical conductivity for EuFe$_2$As$_2$. Furthermore, only one
weak phonon mode around 262 \cm is seen in EuFe$_2$As$_2$,
indicating screening from the conducting carriers is much better
in the 122 system.

The discovery of superconductivity in hole doped ternary iron
arsenide (Ba,K)Fe$_2$As$_2$\cite{Rotter2} immediately attracts
considerable interest in this 122 system.\cite{Krellner, Chen2,
Huang, Zhao, Ni, Chen3} Benefited from high-quality single
crystals, optical studies on the 122-type compounds achieve more
profound progress than that in the 1111 system. A detailed optical
study on single crystal samples of BaFe$_2$As$_2$ and
SrFe$_2$As$_2$ parent compounds was reported by Hu \emph{et
al.}\cite{Hu122}. In this report, a double-gap character was found
for both compounds, along with a dramatic reduction for the
free-carrier spectral weight and scattering rate below T$_{SDW}$.

\subsubsection{SDW double-gap}\label{122SDWgap}

\begin{figure}[t]
\begin{center}
\includegraphics[width=2.7in]{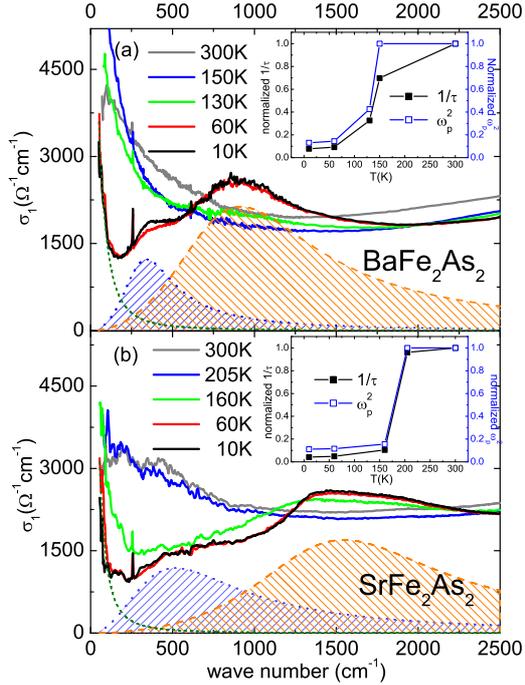}%
\vspace*{-0.20cm}%
\caption{\label{fig:Ba122lowfreq} The real part of conductivity
$\sigma_1(\omega)$ for (a) BaFe$_2$As$_2$ and (b) SrFe$_2$As$_2$
below 2500 \cm. The Drude term (short dash green line) and the
first two Lorentz peaks abstracted from a Drude-Lorentz fit for
T=10 K is shown at the bottom. Inset: normalized 1/$\tau$ (black)
and $\omega_p^2$ (blue).\cite{Hu122}}
\end{center}
\end{figure}

Figure \ref{fig:Ba122lowfreq} demonstrates the low frequency
optical conductivity for BaFe$_2$As$_2$ (T$_{SDW}$=138 K) and
SrFe$_2$As$_2$ (T$_{SDW}$=200 K). The Drude-like conductivity is
seen at high temperature. However, a severe suppression of the
Drude component is seen at low temperature, and the lost spectral
weight gradually piles up into the absorption peaks at higher
energies, suggesting a large amount of density of states are
removed from E$_F$ by the energy gap for T$<$T$_{SDW}$.
Interestingly, a double-peak structure is seen in the conductivity
spectra. Considering the multi-band property for
AFe$_2$As$_2$,\cite{SinghBa122} the double peaks in
$\sigma_1(\omega)$ should correspond to two SDW gaps on separated
Fermi surfaces. Here we note the absorption peaks in
$\sigma_1(\omega)$ develop just below T$_{SDW}$, and the gap-like
peaks no longer exist in the nonmagnetic state. Even for 205 K,
which is just 5 K above the T$_{SDW}$ in SrFe$_2$As$_2$, no gap
signature can be found. Therefore, the fluctuation effect for the
SDW should be rather weak in AFe$_2$As$_2$. From the gap values
and SDW transition temperatures, we obtained the ratio of
2$\Delta$/$k_BT_{SDW}\approx$3.5-3.6 for the smaller gap, and
9-9.6 for the larger gap for the two compounds. The smaller gap
coincides roughly with the gap value expected by the conventional
BCS relation, while the large one is very different.

Because of the presence of two different gap values in the SDW
ordered state, direct information on where the FS sheets are
gapped is highly desired. Several angle resolved photoemission
spectroscopy (ARPES) investigations on AFe$_2$As$_2$ (A=Ba or Sr)
single crystals have been reported, however, two earlier works by
Yang \emph{et al.}\cite{Feng} and Liu \emph{et al.}\cite{Kaminski}
reported complete absence of SDW gaps, in sharp contrast to the
infrared spectroscopy result. Very recently, Hsieh \emph{et al.}
report an orbital resolved ARPES study on SrFe$_2$As$_2$, and find
evidence for an anisotropic SDW gap with energy scales on the
order of 50 meV, which thus tends to be consistent with optical
results.\cite{Hsieh}

\subsubsection{Free-carrier response}\label{122SDWlow}

It is well-known that the low-$\omega$ Drude component comes from
the itinerant carrier contribution. The Drude spectral weight
determines the $\omega_p^2$ ($\omega_p$ is the plasma frequency),
which is proportional to \emph{n/m$_{eff}$} (where \emph{n} is the
carrier density, \emph{m$_{eff}$} is the effective mass); while
its width reflects the carrier scattering rate 1/$\tau$. As shown
in Fig. \ref{fig:Ba122lowfreq}, Drude components are present not
only at T$>$T$_{SDW}$, but also in the SDW state below the
gap-induced absorption peaks. Therefore, the Fermi surface of
AFe$_2$As$_2$ is only partially gapped below T$_{SDW}$. Using a
Drude-Lorentz fit,\cite{Dressel} we can extract the free-carrier
contribution from $\sigma_1(\omega)$. The room \emph{T} plasma
frequency is 12900 \cm for BaFe$_2$As$_2$, and 13840 \cm for
SrFe$_2$As$_2$. Although the plasma frequency is larger than 1.5
eV, one should also note the scattering rate for the normal state
is rather large, i.e., 700 \cm for BaFe$_2$As$_2$ and 950 \cm for
SrFe$_2$As$_2$. Such a large scattering rate corresponds to an
overdamped plasma edge in R($\omega$).\cite{Hu122} Below the SDW
transition, both the plasma frequency and the scattering rate are
dramatically reduced. The variations of 1/$\tau$ and $\omega_p^2$
with temperature for the Drude term are shown in the insets of
Fig. \ref{fig:Ba122lowfreq}. Both parameters are normalized to
their 300 K value. Provided the effective mass of itinerant
carriers does not change with temperature, then the residual
carrier density is only 12$\%$ of that at high temperature for
both compounds. This means that roughly 88$\%$ of FS is removed by
the gapping associated with SDW transitions. On the other hand,
the scattering rate was reduced by about 92-96$\%$. Therefore, the
opening of the SDW partial gap strongly reduces the scattering
channel, leading to a metallic behavior with enhanced dc
conductivity in the gapped state. In the quantum oscillation
measurement on SrFe$_2$As$_2$ single crystal, a large reduction of
the paramagnetic Fermi surface by SDW transition is also observed,
then the remaining Fermi surface pockets are only 2\% of the
original Brillouin zone area in the low \emph{T}
phase.\cite{Sebastian}

It should be remarked that the driving mechanism for the SDW
instability in the parent compound is currently under debate.
Although the stripe- or collinear-type antiferromagnetic order in
the SDW state was first suggested to result from the nesting
between the hole and electron Fermi surfaces (FS) of itinerant
electrons,\cite{DongEPL} it was alternatively proposed that the
superexchange interaction of Fe ions mediated through the
off-plane As atom plays a key role in the spin configuration
formation.\cite{Yildirim,Ma,Fang,Xu,Si,Wu} A stripe-type AFM would
arise when the superexchange interaction between
next-nearest-neighbor Fe sites becomes larger than half of the
nearest neighbor exchange interaction. The optical study of Hu et
al. shed light on this debating issue. Their data clearly
demonstrate that the parent compounds are metallic both above and
below SDW ordering temperatures. They have relatively high plasma
frequency (a bit higher than 1.5 eV) in the normal phase.
Furthermore, the partial gap values below SDW ordering
temperatures are consistent with the expectation of nesting
scenario where the temperature dependence of the gap should
resemble that of BCS theory. Therefore, Hu's work further favors
the itinerant picture for the SDW instability.\cite{DongEPL} Under
such an itinerant picture, superconductivity in the doped system
is mediated by spin fluctuation,\cite{Mazin,Cvetkovic} where a
sign reversal of the order parameters between different Fermi
surfaces is obtained.

\subsubsection{High energy feature}\label{122SDWmid}

\begin{figure}[t]
\begin{center}
\includegraphics[width=2.7in]{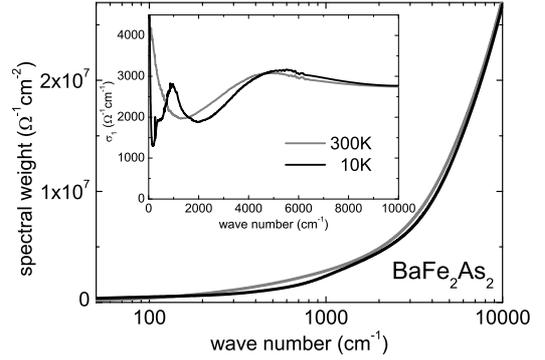}%
\vspace*{-0.20cm}%
\caption{\label{fig:Ba122Escale} The spectral weight for
BaFe$_2$As$_2$ below 10000 \cm. Inset: $\sigma_1(\omega)$ below
10000 \cm (together with the low-$\omega$ extrapolation based on
the \emph{dc} conductivity).\cite{Hu122}}
\end{center}
\end{figure}

As mentioned above, Boris \emph{et al.} observe a high energy
pseudogap around 0.65 eV for LaFeAsO and LaFeAsO$_{0.9}$F$_{0.1}$
(Fig. \ref{fig:BorisPG}), here similar \emph{T}-dependent feature
around 5000 \cm (0.62 eV) also exists in BaFe$_2$As$_2$ and
SrFe$_2$As$_2$. Figure \ref{fig:Ba122Escale} is the spectral
weight analysis for BaFe$_2$As$_2$, from which two characteristic
energy scales can be found: the first one below 2000 \cm and the
second one above 2000 \cm up to 10000 \cm. In the first range
($\omega$$\leq$2000 \cm), the narrowing of the Drude peak and the
development of the SDW double-gap in the mid-infrared lead to a
spectral weight redistribution from the free-carrier contribution
into the gap induced absorption peaks as decreasing \emph{T} (see
the conductivity data in Fig. \ref{fig:Ba122lowfreq}a for
comparison). This redistribution process is not complete until
2000 \cm, which is an energy scale related with the SDW gap. In
the second range (2000-10000 \cm), there is another
\emph{T}-dependent spectral weight transfer from low to high
energies due to the 5000 \cm peak in $\sigma_1(\omega)$. This
feature does not directly related with T$_{SDW}$ as it exists even
in the non-magnetic state.\cite{Hu122} As shown in the inset of
Fig. \ref{fig:Ba122Escale}, the 5000 \cm mid-infrared peak
dominates the conductivity spectrum for 300 K, which slightly
moves towards higher energies for \emph{T}=10 K. Correspondingly,
the low frequency spectral weight also shifts towards higher
frequencies at low \emph{T}, and the total spectral weight
recovers to normal-state value near 9000 \cm (Fig.
\ref{fig:Ba122Escale}).

Similar mid-infrared component has been observed in the doped
sample, e.g., (Ba,K)Fe$_2$As$_2$\cite{BaKPRL} and
Ba(Fe,Co)$_2$As$_2$, and the temperature evolution for this 5000
\cm peak seems unchanged in these doped systems. Such a high
energy feature suggests the incoherent part for Fe 3\emph{d}
bands. Strictly speaking, one can not tell whether it is an
interband transition or a pseudogap based on current optical data.
Anyhow, here we note that a strong band renormalization in the
122-type compounds is found from recent ARPES
studies.\cite{Hsieh,HYLiu,DingBaK,Sato,Sekiba,Terashima}
Therefore, the energy scales for possible interband transitions
calculated from the LDA (local-density approximation) might
deviate from experimental observation. The origin of this
mid-infrared feature requires further studies.

\subsection{Absence of the SDW gap for Fe$_{1+y}$Te} \label{FeTe}

\begin{figure}[t]
\begin{center}
\includegraphics[width=3.5in]{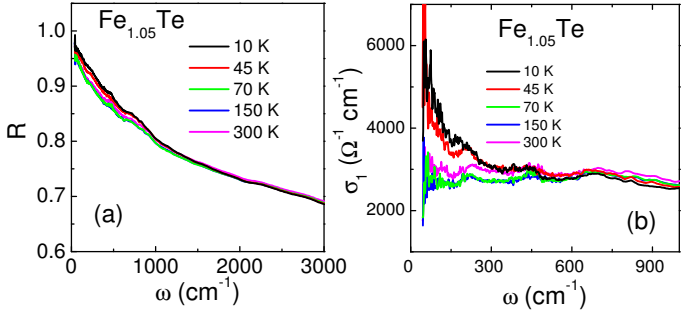}%
\vspace*{-0.20cm}%
\caption{\label{fig:FeTe} (a) Reflectivity R($\omega$) and the
real part of conductivity $\sigma_1(\omega)$ for
Fe$_{1.05}$Te.\cite{ChenFeTe}}
\end{center}
\end{figure}

As shown above, the SDW gap induced suppression can be found for
both ReFeAsO and AFe$_2$As$_2$. However, the 11-type single
crystalline Fe$_{1.05}$Te, which undergoes a structural distortion
along with the establishment of a long range antiferromagnetic
order near 65K,\cite{Bao,Li} shows no clear signature of the SDW
gap.

Superconductivity with transition temperature T$_c$ up to 15 K was
obtained on Fe$_{1+x}$(Se,Te) system at ambient
pressure.\cite{Wu01,Mao,Wu02} The T$_c$ can go up to 27 K at a
pressure of 1.48 GPa.\cite{Mizuguchi} Figure \ref{fig:FeTe} shows
the reflectivity R($\omega$) and the real part of conductivity
$\sigma_1(\omega)$ for Fe$_{1.05}$Te.\cite{ChenFeTe} The normal
state reflectivity is similar to AFe$_2$As$_2$,\cite{Hu122} that
R($\omega$) has an almost linear frequency dependence up to
mid-infrared region. It is important to note that the conductivity
spectra above T$_{SDW}$ are rather flat. Definitely, it is not a
semiconductor as no semiconductor-like gap could be found in
$\sigma_1(\omega)$, but it is not a simple metal as well, because
of the absence of a Drude-like peak. As is well known, the width
of Drude peak is determined by the scattering rate (or inverse of
the transport lifetime) of the quasiparticle, the measurement
result indicates that there is no well defined quasiparticle with
sufficiently long transport life time above T$_{SDW}$. The charge
transport is rather incoherent. Below T$_{SDW}$, the low-$\omega$
R($\omega$) increases fast. Correspondingly, a Drude component in
$\sigma_1(\omega$) develops quickly from the incoherent
background. Unlike the parent compounds of LaFeAsO and
AFe$_2$As$_2$ (A=Ba, Sr), there is no partial gap formation in
optical spectra below T$_{SDW}$.\cite{ChenFeTe} The absence of SDW
gap formation has been confirmed by recent ARPES experiment on
Fe$_{1.05}$Te.\cite{XiaFeTe} Those data, together with transport
and specific heat measurement results, suggest that
Fe$_{1+x}$(Se,Te) system is an exceptional case. This was
presumably attributed to the presence of excess Fe$^{1+}$ ions,
which have large magnetic moments and locate randomly in the Fe(2)
sites as in Fe$_2$As compound, resulting in rather strong
interaction with the magnetism of FeTe
layers.\cite{JZhang,ChenFeTe}

\section{Superconductivity in the doped compound} \label{SCgap}
Probing the energy scale and the symmetry of the superconducting
gap are always important tasks for infrared spectroscopy. In the
early stage, when only 1111-type polycrystals are available, the
superconducting gap was observed as an abrupt increase of
reflectivity in LaFeAsO$_{1-x}$F$_x$.\cite{ChenPRLLa} Soon, the
superconducting gap in (Nd,Sm)FeAsO$_{0.82}$F$_{0.18}$
polycrystals were found by infrared ellipsometry.\cite{Dubroka} In
addition, a pronounced dip at 20 meV was found in the reflectivity
ratio R($\omega$,T$<$\Tc)/R($\omega$,T$_n$) from a series F-doped
SmO$_{1-x}$F$_x$FeAs polycrystals with unpolished
surfaces.\cite{Mirzaei} The electron-boson coupling was estimated
by Drechsler \emph{et al.} based on optical data collected on
powder LaFeAsO$_{0.9}$F$_{0.1}$ and reported value of in-plane
penetration depth.\cite{DrechslerLa} With the availability of
single crystalline samples of (A,K)Fe$_2$As$_2$ (A=Sr, Ba), more
reliable in-plane optical data were obtained. The first optical
evidence for an s-wave superconducting gap was found in
Ba$_{0.6}$K$_{0.4}$Fe$_2$As$_2$,\cite{BaKPRL} where the in-plane
R($\omega$) suddenly turns up and approaches unity below \Tc,
following the well-established BCS description.

\subsection{Probing the superconducting gap} \label{SC}
Three length scales characterize the electrodynamics of
superconductors, i.e., the London penetration depth $\lambda_L$,
the correlation length $\xi_0$, and the mean free path of the
uncondensed electrons $l$. The London penetration depth
$\lambda_L$=$c/\omega_{ps}$, where $\omega_{ps}$ is the plasma
frequency for the condensed carriers. The penetration depth is a
characteristic scale within which the external magnetic field is
exponentially screened. The correlation length $\xi_0$ is the
spatial extension of the Cooper pairs.\cite{Tinkham, Dressel} For
superconductors, the dirty limit is the case in which $l\ll\xi_0$,
thus the width of the Drude term 1/$\tau$ in the normal state
should be larger than the superconducting gap 2$\Delta$; while the
clean limit is a case that $l\gg\xi_0$, thus
1/$\tau$$\ll$2$\Delta$. The dirty limit and the clean limit can be
distinguished by their different spectral responses across the
superconducting transition. In the clean limit, no clear signature
of the superconducting gap could be found in $\sigma_1(\omega)$,
because the spectral weight for the superconducting condensate
lies below 2$\Delta$, so that there is no detectable change of
spectrum at the gap energy 2$\Delta$ across the T$_c$. In the
dirty limit, the normal-state Drude term is broader than
2$\Delta$, therefore the disappearance of single-particle
excitation below 2$\Delta$ would left an incoherent contribution
in $\sigma_1(\omega)$, and the normalized spectral weight of the
superconducting condensate converges much more slowly than in the
clean-limit case. \cite{Homes05, Homes04}

According to the optical sum rule\cite{Dressel}
$\int\sigma_1(\omega)d\omega$=$\omega_p^2$/8, the area under
$\sigma_1(\omega)$ should be conserved, and the total spectral
weight is determined by the effective carrier density as the
plasma frequency $\omega_p^2$=4$\pi N e^2/m_{eff}$. In this sense,
the lost spectral weight in $\sigma_1(\omega)$ below \Tc
(so-called "missing area") is a more essential character than the
gap in optical spectrum.\cite{Tinkham} The "missing area" denotes
the existence of superconducting condensate, i.e., a collective
excitation at zero frequency with the form of $\delta(\omega)$
function. Using Kramers-Kronig relation, a term $\sigma_1(\omega)$
= $A\delta(\omega)$ corresponds to $\sigma_2(\omega)$ = 2$A/\pi
\omega$. Comparing with the London equation $\sigma_2(\omega)$ =
$c^2$/4$\pi \lambda_L^2 \omega$, the penetration depth $\lambda$
calculated from the imaginary part of conductivity should be equal
to that calculated from the "missing area" $A$ in the real part of
conductivity,\cite{Tinkham}
\begin{equation}
\lambda = c/\sqrt{8A} \label{eq:MissingArea}
\end{equation}
which is the famous Ferrell-Glover-Tinkham (FGT) sum
rule.\cite{Ferrell58, Tinkham59} As the imaginary part of the
conductivity $\sigma_2(\omega)$ diverges as 1/$\omega$ below
2$\Delta$, it can be shown that the optical reflectivity
R($\omega$) would rapidly approach to unity below the gap energy
for BCS superconductor.\cite{Dressel} Therefore, the up-turn in
R($\omega$) at low frequency for \emph{T}$<$\Tc is an optical
evidence for the superconducting gap, and a rapid saturation of
reflectivity indicates the gap is isotropic (s-wave). While for
the case of non s-wave pairing, the optical reflectivity will
slowly increases towards unity instead of an abrupt jump.

\subsection{Superconducting gap in 1111-type polycrystal} \label{SCin1111}

\begin{figure}[t]
\begin{center}
\includegraphics[width=2.7in]{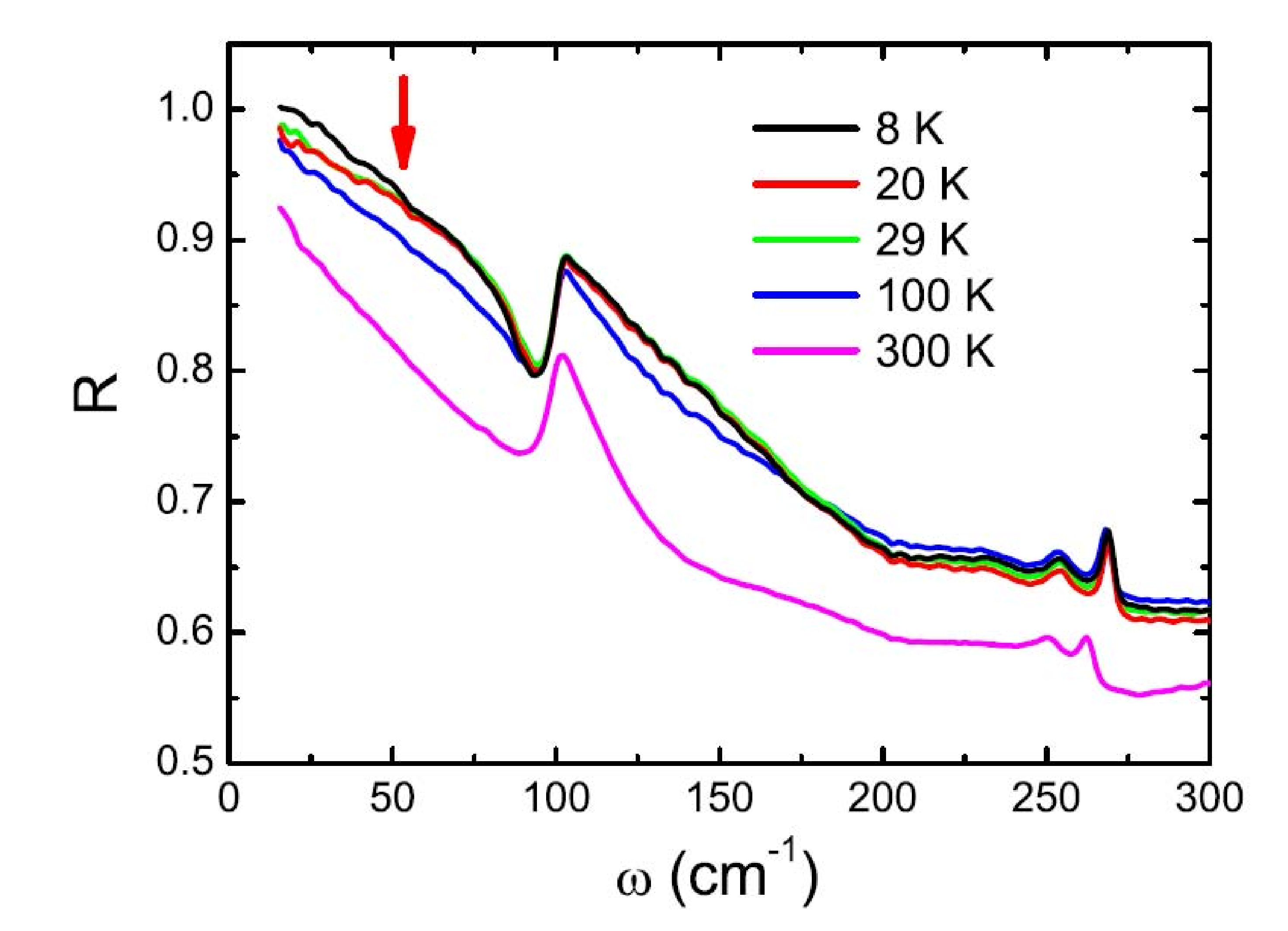}%
\vspace*{-0.20cm}%
\caption{\label{fig:LaFeAsOF} The optical reflectivity R($\omega$)
for F-doped LaFeAsO polycrystal.\cite{ChenPRLLa}}
\end{center}
\end{figure}

The first optical investigation for the superconducting gap in
FeAs-based superconductor was done on a LaFeAsO$_{1-x}$F$_x$
polycrystal. Transport measurements testify the existence of
superconducting transition, and the \emph{ac} magnetic
susceptibility drops at 20 K.\cite{ChenPRLLa} Although the
spectral change from 29 K to 20 K is hard to resolve in Fig.
\ref{fig:LaFeAsOF}, an evident up-turn for T=8 K below 50$\sim$60
\cm can be found (see the arrow in Fig. \ref{fig:LaFeAsOF}). Since
the up-turn in optical reflectivity R($\omega$) corresponds to the
onset energy of the superconducting gap, then 2$\Delta$/$k T_c$ is
3.5-4.2 for LaFeAsO$_{1-x}$F$_x$ polycrystal, which is quite close
to the prediction by BCS theory. Unlike the conventional BCS-type
superconductor, here the reflectivity for LaFeAsO$_{1-x}$F$_x$
does not abruptly approach unity below 2$\Delta$, but gradually
increases towards the 100\% line. A continuous increase in
R($\omega$) is a common feature for high-\Tc cuprate where the
superconducting gap has a d-wave symmetry.\cite{Homes06} However,
no conclusion on the gap symmetry could be drawn from optical data
on polycrystalline sample.

\begin{figure}[t]
\begin{center}
\includegraphics[width=2.7in]{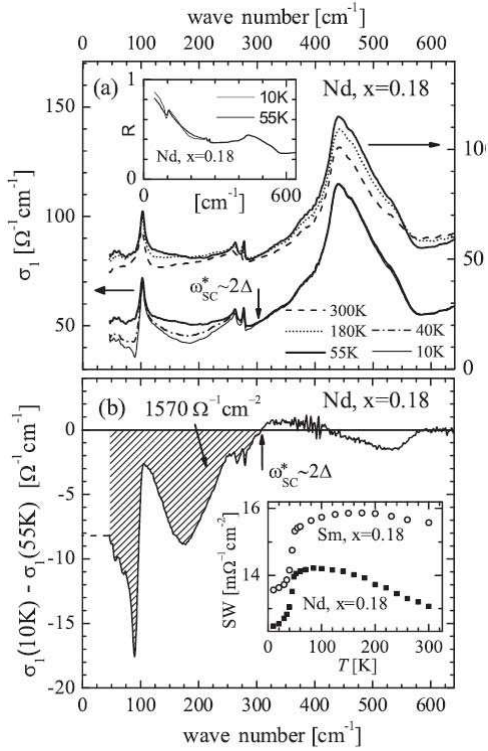}%
\vspace*{-0.20cm}%
\caption{\label{fig:Dubroka} The optical response for
NdFeAsO$_{0.82}$F$_{0.18}$ (Nd, x=0.18) and
SmFeAsO$_{0.82}$F$_{0.18}$ (Sm, x=0.18) superconductors. (a)
$\sigma_1(\omega)$ of the Nd, x=0.18 sample. The arrow marks the
onset of the superconductivity-induced gaplike suppression. Inset:
calculated reflectivity. (b) $\sigma_1(10 K)$-$\sigma_1(55 K)$.
The shaded area is the missing spectral weight due to the
superconducting condensate. Inset: integrated $\sigma_1(\omega)$
(i.e., the spectral weight) between 45 and 300 \cm.\cite{Dubroka}}
\end{center}
\end{figure}

Further optical evidence for the superconducting gap in Fe-based
superconductor is obtained from ellipsometric spectroscopy on
polycrystalline NdFeAsO$_{0.82}$F$_{0.18}$ (T$_c$=52 K) and
SmFeAsO$_{0.82}$F$_{0.18}$ (T$_c$=45 K).\cite{Dubroka} Figure
\ref{fig:Dubroka}a shows the optical conductivity for
NdFeAsO$_{0.82}$F$_{0.18}$ polycrystal at selected temperatures.
The optical conductivity is dominated by four pronounced phonon
modes in the far-infrared region for all temperatures. Regardless
of these phonon modes, one can find the \emph{T}-evolution for
$\sigma_1(\omega)$ is changed across \Tc: the $\sigma_1(\omega)$
is slightly enhanced due to the metallic response above \Tc; while
an obvious gap-like suppression below 300 \cm could be found for
T$<$\Tc. The suppression can be more clearly resolved by the
"missing area" in $\sigma_1(\omega)$. Figure \ref{fig:Dubroka}b
shows the difference spectrum $\sigma_1(10 K)$-$\sigma_1(55 K)$ as
a function of frequency, and the inset plots the
\emph{T}-dependent spectral weight between 50 an 300 \cm. Dubroka
\emph{et al.} defines the superconducting gap 2$\Delta$ as the
onset energy for the optical suppression $\omega_{SC}^*$ (300
\cm$\sim$37 meV), thus 2$\Delta$/$k_B T_c$$\approx$8.

As shown above, optical signatures of the superconducting gap are
found in polycrystalline LaFeAsO$_{1-x}$F$_x$,
SmFeAsO$_{1-x}$F$_x$, and NdFeAsO$_{1-x}$F$_x$. However, both
methods used in Fig. \ref{fig:LaFeAsOF} and \ref{fig:Dubroka} to
identify the gap onset energy are not accurate. The pronounced
phonon mode near 100 \cm strongly affects the low frequency
reflectivity, thus it is difficult to distinguish the exact
frequency for the up-turn in R($\omega$). While subtracting the
normal state spectra to cancel out the phonon information and
using $\omega_{SC}^*$ (Fig. \ref{fig:Dubroka}) as the gap will
overestimate the gap in the dirty-limit.\cite{Homes05,Homes04}
Then spectroscopic data on the single crystals are highly
desirable.

\subsection{Superconducting gap in 122-type single crystal} \label{SCin122}
The first optical data on superconducting
Ba$_{1-x}$K$_x$Fe$_2$As$_2$ single crystals were reported
individually by two groups.\cite{BaKPRL, YangBaK} Li \emph{et al.}
gave evidence for an s-wave superconducting gap in optimally doped
Ba$_{0.6}$K$_{0.4}$Fe$_2$As$_2$, and found the
Ferrell-Glover-Tinkham sum rule was satisfied within an energy
scale of 6$\Delta$.\cite{BaKPRL} Yang \emph{et al.} studied the
optical properties of Ba$_{0.55}$K$_{0.45}$Fe$_2$As$_2$ above its
superconducting transition temperature (28 K), and concentrated on
the normal state properties.\cite{YangBaK}

\begin{figure}[t]
\begin{center}
\includegraphics[width=2.7in]{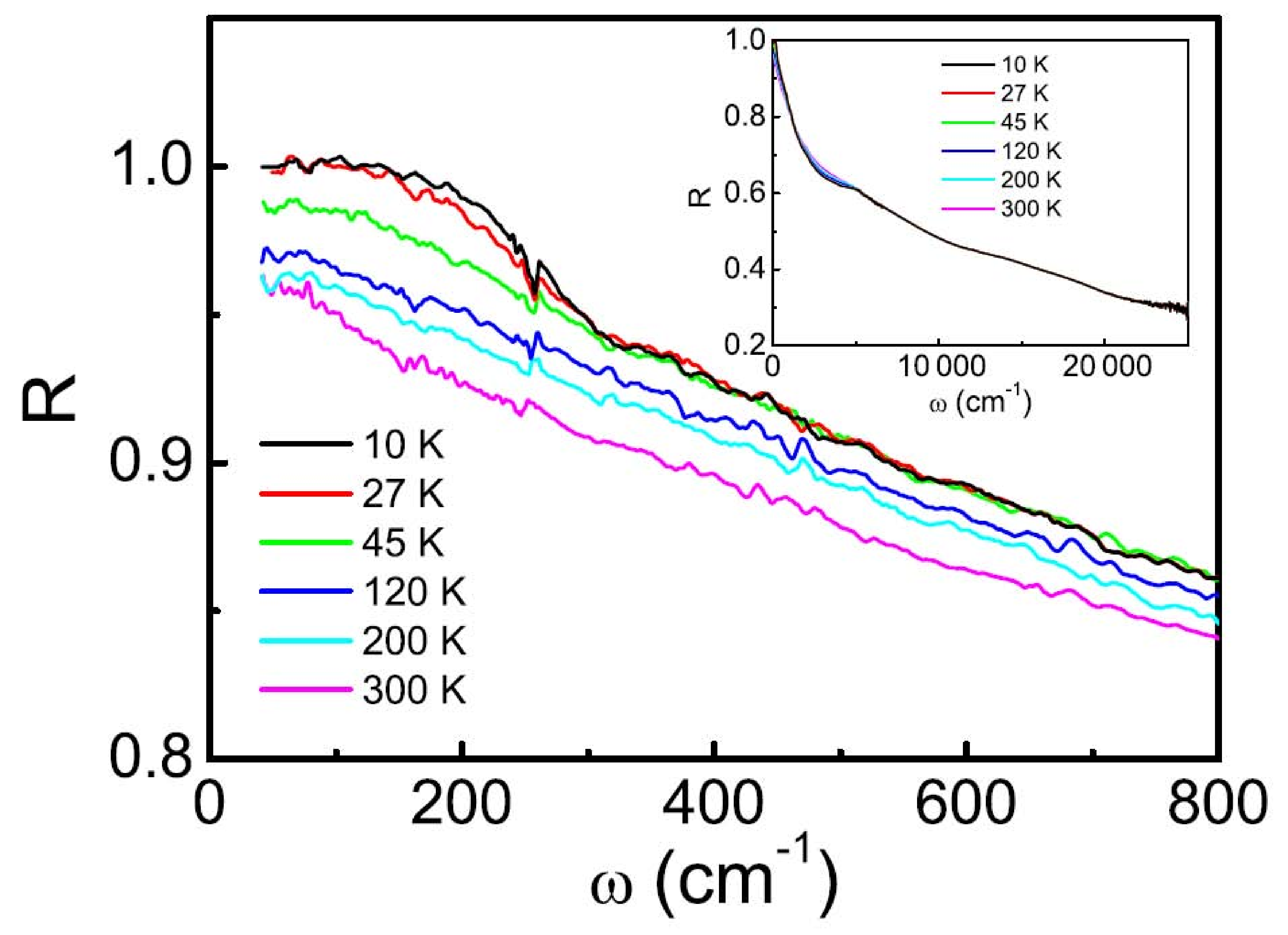}%
\vspace*{-0.20cm}%
\caption{\label{fig:BaKReflectivity} Optical reflectivity
R($\omega$)for Ba$_{0.6}$K$_{0.4}$Fe$_2$As$_2$ in the far-infrared
region. Inset: The inset shows R($\omega$) over a broad frequency
range.\cite{BaKPRL}}
\end{center}
\end{figure}

\begin{figure}[b]
\begin{center}
\includegraphics[width=2.7in]{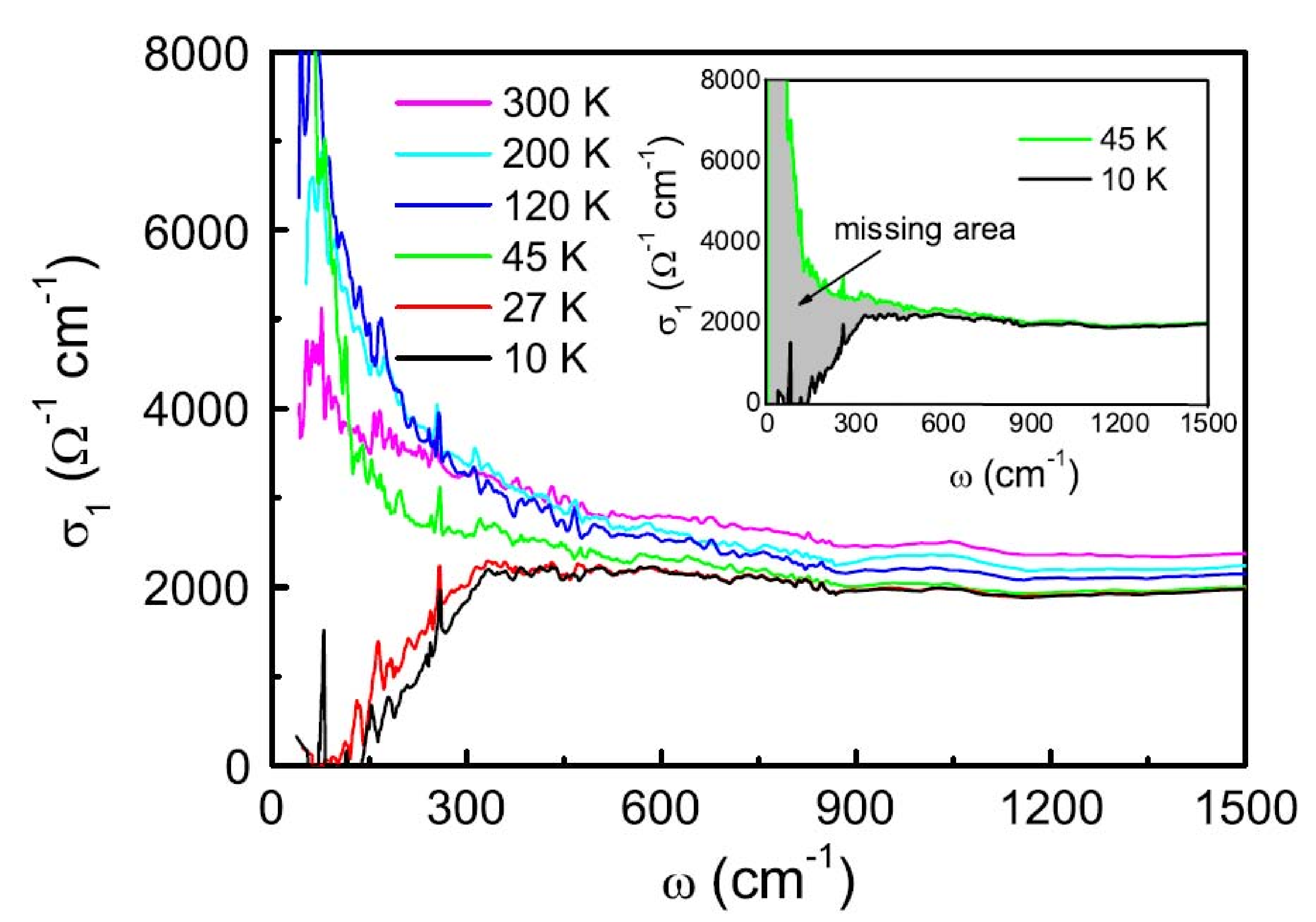}%
\vspace*{-0.20cm}%
\caption{\label{fig:BaKSigma} Optical conductivity for
Ba$_{0.6}$K$_{0.4}$Fe$_2$As$_2$. The inset shows
$\sigma_1(\omega)$ at 10 and 45 K. The shaded area represents the
"missing area" due to the opening of superconducting energy
gap.\cite{BaKPRL}}
\end{center}
\end{figure}

Figure \ref{fig:BaKReflectivity} shows the ab-plane optical
reflectivity for Ba$_{0.6}$K$_{0.4}$Fe$_2$As$_2$ (\Tc=37 K). The
R($\omega$) for all T$>$\Tc follow a linear frequency dependence.
When T$<$\Tc, the R($\omega$) suddenly turns up below 300 \cm and
approaches unity around 150 \cm, which is a typical character for
an s-wave superconducting gap.\cite{Dressel} The temperature
dependence of the optical conductivity is shown in Fig.
\ref{fig:BaKSigma}. The Drude component narrows as decreasing
\emph{T} from 300 K to 45 K. For T$<$\Tc, a large amount of the
Drude weight in $\sigma_1(\omega)$ collapse into the
superconducting condensate at zero frequency, leading to a
non-zero excitation above 2$\Delta$, and a "missing area" between
10 K and 45 K (see the inset figure). As shown in Section
\ref{SC}, the "missing area" in $\sigma_1(\omega)$ corresponds to
the zero-frequency collective mode. The penetration depth
estimated from the missing area is 2080 $\AA$. The same quantity
calculated from the imaginary part of optical conductivity
$\sigma_2(\omega)$=$c^2$/4$\pi \lambda^2 \omega$ is
$\lambda$$\simeq$1950 $\AA$. The good agreement for the
penetration depth by different calculations indicated the validity
of FGT sum rule in Ba$_{0.6}$K$_{0.4}$Fe$_2$As$_2$. An inspection
of the inset of Fig. \ref{fig:BaKSigma} reveals that the "missing
area" extends to the frequency roughly below 600 \cm, about 3
times larger than the higher superconducting energy gap 2$\Delta$.
This indicates that the superconducting condensate forms rapidly
or the FGT sum rule is rapidly recovered. This is very different
from underdoped high-T$_c$ cuprates where recovery of the FGT sum
rule goes to very high energy, or the FGT sum rule is even
violated.\cite{Homes05,Molegraaf,Bontemps}

\begin{figure}[t]
\begin{center}
\includegraphics[width=2.7in]{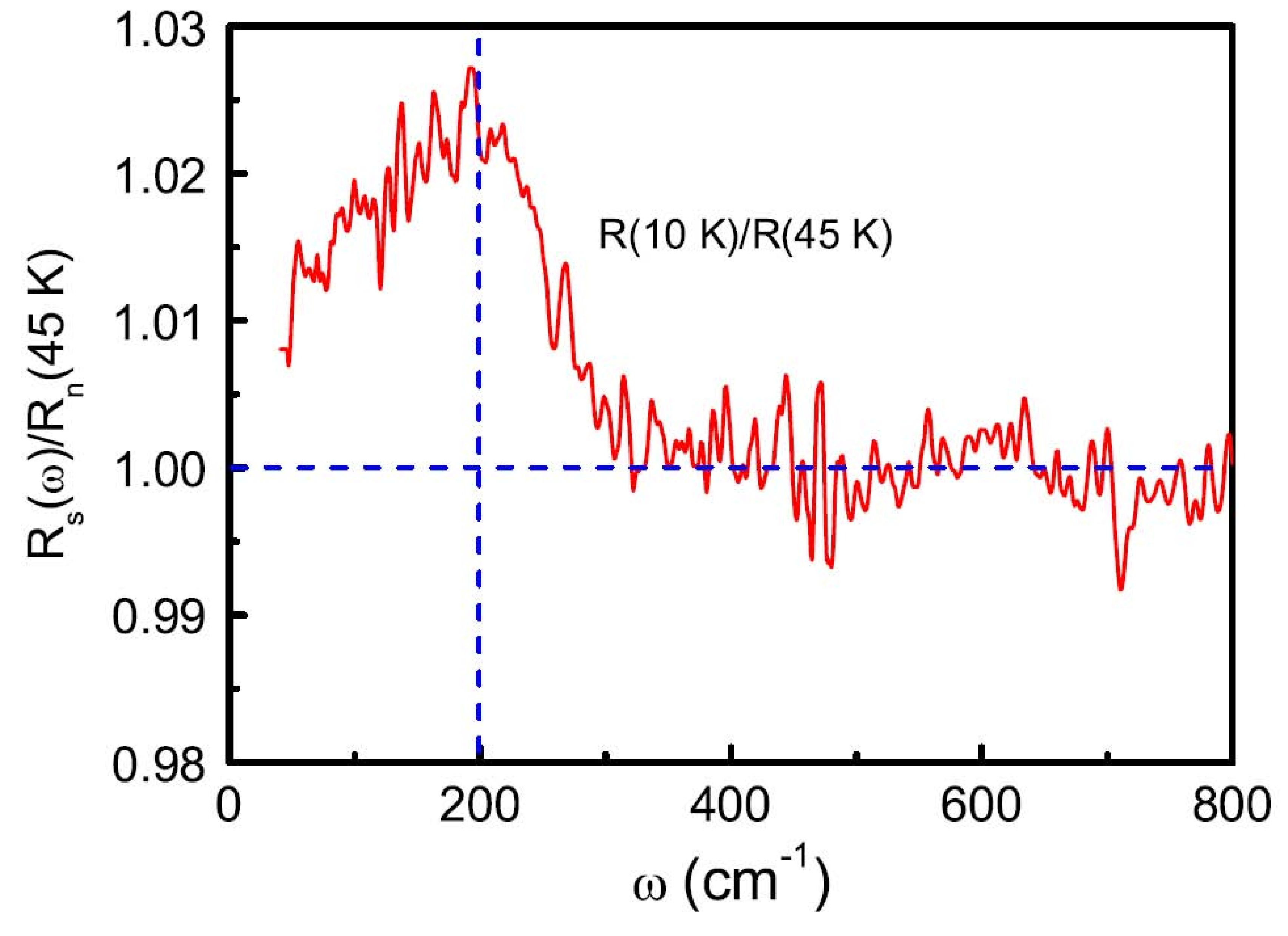}%
\vspace*{-0.20cm}%
\caption{\label{fig:BaKR} The reflectance ratio R($\omega$, T=10
K)/R($\omega$, T=45 K). A peak near 200 \cm is seen.\cite{BaKPRL}}
\end{center}
\end{figure}

According to the BCS theory in the dirty limit, there is no
optical excitation for $\hbar \omega$$<$2$\Delta$, while above
2$\Delta$ the optical conductivity $\sigma_1(\omega)$ gradually
rises due to the case II coherence factor.\cite{Tinkham} Since the
$\sigma_1(\omega)$ for $\omega<$2$\Delta$ is condensed into the
zero frequency collective mode, $\sigma_2(\omega)$ will diverge as
1/$\omega$ for $\omega$$<$2$\Delta$, then a peak at 2$\Delta$
could be found in
R($\omega$,T$\ll$\Tc)/R($\omega$,T$\simeq$\Tc).\cite{Dressel} As
shown in Fig. \ref{fig:BaKSigma}, the conductivity is almost zero
below roughly 150 \cm due to the flat and close to unity
R($\omega$), which is an optical evidence for the s-wave
superconducting energy gap. However, the peak in R($\omega$,10
K)/R($\omega$,45 K) is around 200 \cm, thus yields a different
energy scale for the superconducting gap. According to recent
ARPES study on Ba$_{0.6}$K$_{0.4}$Fe$_2$As$_2$ (where the single
crystalline sample comes from the same batch as the one used in
optical study\cite{BaKPRL}), two isotropic gaps
($\Delta_1$=6$\sim$8 meV, $\Delta_2$=10$\sim$12 meV) are found on
different Fermi surfaces.\cite{DingARPES,ZhaoARPES,Nakayama} Thus
the discrepancy in the gap energies from $\sigma_1(\omega)$ and
R$_s$/R$_n$ might also suggest a two-gap character for
Ba$_{0.6}$K$_{0.4}$Fe$_2$As$_2$.

\subsection{Coherence factor and the double-gap in Ba$_{0.6}$K$_{0.4}$Fe$_2$As$_2$} \label{BCScoherencefactor}
\begin{figure}[t]
\begin{center}
\includegraphics[width=2.7in]{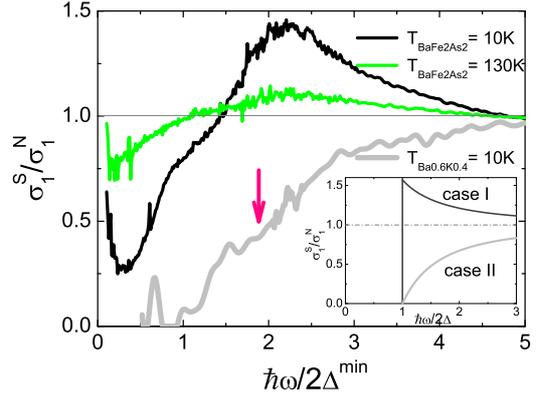}%
\vspace*{-0.20cm}%
\caption{\label{fig:CF} $\sigma_{1}^S$/$\sigma_{1}^N$ for
BaFe$_2$As$_2$ and Ba$_{0.6}$K$_{0.4}$Fe$_2$As$_2$. The
double-peak character reflects a double-gap feature for the SDW
state in BaFe$_2$As$_2$. Superconducting
Ba$_{0.6}$K$_{0.4}$Fe$_2$As$_2$ also shows a weak double-gap
feature (the absorption edge for the second gap is marked by
arrow). Inset: optical conductivity calculated by Eq.\ref{eq:CF2}
(one isotropic gap) for case I and case II coherence factors at T
=0.}
\end{center}
\end{figure}

Single electron transitions, induced by external perturbations
such as electromagnetic radiation, ultrasound, or nuclear
relaxation are determined by coherence factors, in addition to
density of state effects.\cite{Gruner} Considering the transition
rate $\alpha_s$ between energy levels \emph{E} and
\emph{E'}=\emph{E}+$\hbar\omega$,
\begin{equation}
\alpha _s  = \int {\left| M \right|^2 } F(\Delta ,E,E')N_s (E)N_s
(E')[f(E) - f(E')]dE \label{eq:CF1}
\end{equation}
where \emph{M} is the magnitude of the one-electron matrix element
thus $\left| M \right|^2 $ the transition probability, $\Delta$
the single particle gap, \emph{N(E)} the density of states,
\emph{f(E)} the Fermi distribution function $1/(1 + e^{E/k_B T})$,
and \emph{F($\Delta$,E,E')} the coherence factor. As shown by
Tinkham\cite{Tinkham} and Gr\"{u}ner\cite{Gruner}, the
electromagnetic absorption process for the density-wave and BCS
superconducting ground states have different coherence factors
since they are invariant under different symmetry operations.
Thus,
\begin{equation}
\frac{{\sigma _{1}^S }}{{\sigma _{1}^N}} = \frac{1}{{\hbar \omega
}}\int\limits_{ - \infty }^\infty  {\frac{{\left| {E(E + \hbar
\omega ) \mp \Delta ^2 } \right|[f(E) - f(E + \hbar \omega
)]}}{{(E^2  - \Delta ^2 )^{1/2} [(E + \hbar \omega )^2  - \Delta
^2 ]^{1/2} }}dE} \label{eq:CF2}
\end{equation}
where $\sigma_{1}^S$ is the conductivity for the density-wave or
superconducting state, $\sigma_{1}^N$ the conductivity for the
normal state, and the upper sign corresponds to case I (e.g., SDW)
and the lower to case II (e.g., BCS superconductor). Calculation
for the conductivity ratio based on Eq. \ref{eq:CF2} requires
numerical integration. However, the integral has a simple
expression in terms of complete elliptic integrals for \emph{T} =
0 K.\cite{Tinkham} The inset of Fig. \ref{fig:CF} shows the
calculation result based on Eq. \ref{eq:CF2} for the single-gap
case. For the SDW ground state with an isotropic gap (case I),
$\sigma_1^S$/$\sigma_1^N$ shows an asymmetry peak at the gap
energy 2$\Delta$, while for a BCS superconductor (case II)
$\sigma_1^S$/$\sigma_1^N$ increases gradually above 2$\Delta$. As
T/T$_c$ (or T/T$_{SDW}$) appears in the exponent factor
(1/[exp($\Delta$/k$_B$T)+1]) for numerical integration, deviation
from the T=0 K case can be neglected when T$\ll$T$_c$ (or
T$\ll$T$_{SDW}$). The calculation can be further extended into the
multiple-gap case, for instance, two BCS gaps on different Fermi
surfaces. Model calculation for the two-gap case has been done by
Barabash and Stroud.\cite{Stroud}

Figure \ref{fig:CF} shows the normalized optical conductivity
$\sigma_1^S$/$\sigma_1^N$ in the broken symmetry state for two
122-type single crystals (the SDW state for BaFe$_2$As$_2$ parent
compound and the superconducting state for
Ba$_{0.6}$K$_{0.4}$Fe$_2$As$_2$). Different from the single-gap
case as shown in the inset of Fig.\ref{fig:CF}, the optical
conductivity for BaFe$_2$As$_2$ has a double-peak character with
two energy scales ($\Delta_1\equiv2\Delta^{min}$ and
$\Delta_2$$\simeq$2$\Delta_1$). In addition, both peaks in
$\sigma_1^S$/$\sigma_1^N$ for BaFe$_2$As$_2$ deviate from the very
sharp peak from theoretical calculation of the isotropic gap as
shown in the inset figure. Even for the lowest temperature
\emph{T}=10 K, the double peaks are still rather broad and round
in shape, then one can hardly fit these experimental data by two
isotropic gaps with the case I coherence factor. This could be in
part attributed to the very complicated band structure with only
partially gapped Fermi surfaces in the SDW ordered state.

\begin{figure}[t]
\begin{center}
\includegraphics[width=2.7in]{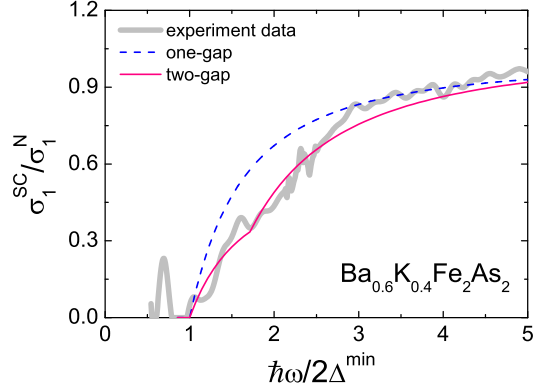}%
\vspace*{-0.20cm}%
\caption{\label{fig:CFforBaK} The optical conductivity ratio
$\sigma_1^{10K}/\sigma_1^{45K}$ compared with the one-gap
($\Delta_1$=7 meV) and two-gap ($\Delta_1$=7 meV and $\Delta_2$=12
meV) fit for the BCS ground state.}
\end{center}
\end{figure}

Similar to the SDW double-gap in the parent compound, the
superconducting Ba$_{0.6}$K$_{0.4}$Fe$_2$As$_2$ also shows a
double-gap character after a normalization process for the optical
conductivity (the onset of the second gap is marked by the arrow).
As shown below, here the symmetry of the superconducting gap is
s-wave. Using a two-gap model ($\Delta_1$=7 meV and $\Delta_2$=12
meV, we can reasonably reproduce the experimental data, while the
calculation for a single gap at 7 meV shows larger discrepancy in
the low frequency region (Fig.\ref{fig:CFforBaK}). Then one could
find that the onset of optical excitation is determined by the
smaller gap, but the fine structure in $\sigma_1(\omega)$ and the
characteristic energy scale in R$_s$/R$_n$ are dependent on the
larger gap. Since the two-component s-wave gap function can well
reproduce the experimental character, the superconducting gaps in
optimally doped (Ba,K)Fe$_2$As$_2$ should be two individual
isotropic gaps on separated Fermi surfaces, which agrees well with
recent ARPES observation.\cite{DingARPES,Nakayama} More
interestingly, ARPES studies find the larger superconducting gap
comes from the $\alpha$ and $\gamma$ Fermi pockets which are well
connected by the SDW vector \textbf{q} in the parent compound,
thus the inter-band interactions might play important role in the
superconducting pairing, as expected also by theoretical
calculations.\cite{Mazin} Further spectroscopic evidence for the
connection between nesting vector and superconducting gap can be
found in another recent ARPES study on
BaFe$_{1.85}$Co$_{0.15}$As$_2$ (an optimally electron-doped 25.5 K
superconductor), in which the $\alpha$ pocket disappears because
of electron doping, and the superconducting gap in $\beta$ pocket
is then comparable with that on $\gamma$ pocket (here $\gamma$ is
well connected with $\beta$ by the vector \textbf{q} after Co
doping).\cite{Terashima}

\section{Electronic Correlation} \label{correlation}
Investigation on Fe-based materials is believed to promote the
research of finding new types of superconductors with higher \Tc.
On the other hand, experimental study on the similarities and
differences between Fe-based superconductors and the high-\Tc
cuprates should help for elucidating the role of magnetic
fluctuations in the superconducting pairing. Optical studies on
the normal state properties for single crystalline
LaFePO\cite{Qazilbash} and
Ba$_{0.55}$K$_{0.45}$Fe$_2$As$_2$\cite{YangBaK} suggest that the
many-body effects cannot be neglected.

\subsection{Electron-boson coupling in LaFePO}\label{LaFePO}
LaFePO is the first superconducting Fe-based compound discovered
by Hosono's group in 2006.\cite{LaFePO} It is also the first
1111-type compound from which relatively large single crystal has
been successfully synthesized.\cite{LaFePOsingle} Unlike the
FeAs-based compounds, LaFePO does not have the antiferromagnetic
transition at high temperature. A recent study on ReFePO (Re=La,
Pr, Nd) single crystals finds a different \Tc evolution that \Tc
decreases as Ln is varied from La to Sm for ReFePO but increases
for ReFeAsO$_{1-x}$F$_x$.\cite{Baumbach} In addition, a linear
\emph{T}-dependence for the penetration depth is found for
LaFePO,\cite{Fletcher} suggesting a different superconducting
mechanism in FeP- and FeAs-based superconductors. However, as
large single crystalline 1111-type FeAs-based superconductor is
still unavailable at present, spectroscopic study on LaFePO is
useful to understanding the normal state properties for the
Fe-based parent compound. Recent ARPES investigation indicates the
ground state for LaFePO is itinerant, with no low-energy kink and
no pseudogap for the bands near E$_F$, thus is clearly
distinguishable from copper oxide superconductors.\cite{Lu}

\begin{figure}[t]
\begin{center}
\includegraphics[width=2.7in]{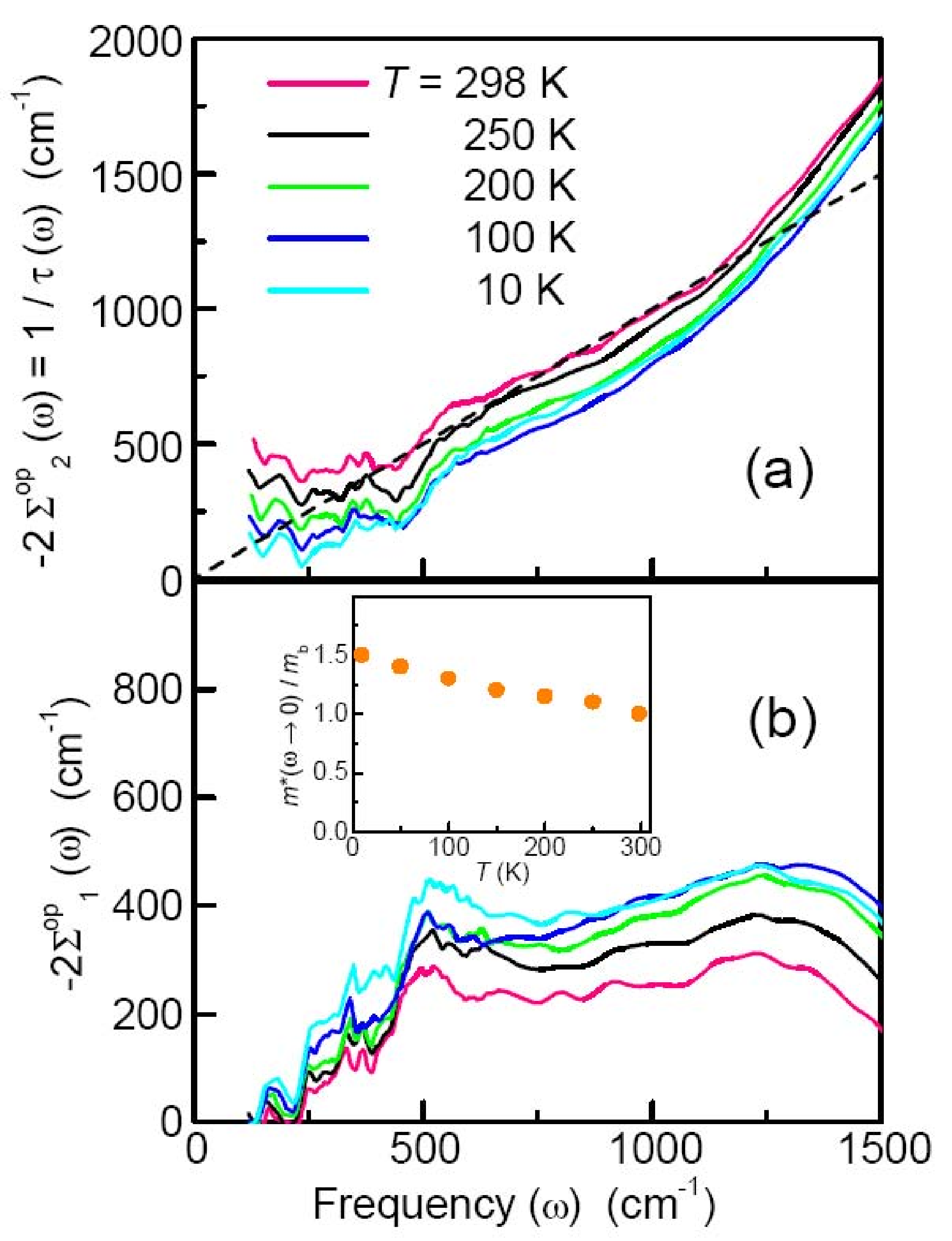}%
\vspace*{-0.20cm}%
\caption{\label{fig:Qazilbash} The frequency dependence of the (a)
imaginary part of the optical quasiparticle self energy
-2$\Sigma_2^{op}(\omega)$ (or scattering rate 1/$\tau(\omega)$),
and (b) the real part of the optical quasiparticle self energy
-2$\Sigma_1^{op}(\omega)$ for LaFePO in the normal state. The
dashed line in panel (a) represents the 1/$\tau$=$\omega$ line.
The inset in panel (b) shows the mass enhancement factor
m$^*$/m$_b$ for $\omega$$\rightarrow$0.\cite{Qazilbash}}
\end{center}
\end{figure}

Although the ARPES study does not find any low-energy kink for
LaFePO single crystal, there is an optical evidence for a 62 meV
electron-boson coupling mode in the optical quasiparticle
self-energy.\cite{Qazilbash} From Qazilbash \emph{et al.}'s data,
the Drude term in $\sigma_1(\omega)$ is well separated from the
high energy interband transitions (see Fig. 2 in their paper
\cite{Qazilbash}), thus an extended Drude analysis is used to
extract the optical quasiparticle self-energy
($\Sigma_1^{op}(\omega)$, $\Sigma_2^{op}(\omega)$) from the real
and imaginary parts of the optical conductivity. Then the
scattering rate 1/$\tau(\omega)$ and mass enhancement factor
$m^*(\omega)/m_b$ can be obtained from the optical quasiparticle
self-energy by
\begin{equation}
 - 2\sum\nolimits_2^{op} {(\omega )}  = \frac{1}{{\tau (\omega )}} = \frac{{\omega _p^2 }}{{4\pi }}(\frac{{\sigma _1^{} (\omega )}}{{\sigma _1^2 (\omega ) + \sigma _2^2 (\omega )}})
\label{eq:SelfE1}
\end{equation}
\begin{equation}
1 - \frac{2}{\omega }\sum\nolimits_1^{op} {(\omega )}  =
\frac{{m^* (\omega )}}{{m_b }} = \frac{{\omega _p^2 }}{{4\pi
}}(\frac{{\sigma _2^{} (\omega )}}{{\sigma _1^2 (\omega ) + \sigma
_2^2 (\omega )}})\label{eq:SelfE2}
\end{equation}
As the Fe-based superconductors are multi-band systems, here the
self-energy should be regarded as an average of the contributions
from the relevant bands.

As shown in Fig. \ref{fig:Qazilbash}a, the scattering rate
1/$\tau(\omega)$ continuously increases with frequency up to 1500
\cm with no saturation. A small kink with weak temperature
dependence could be found around 500 \cm. This kink in
1/$\tau(\omega)$ corresponds to the 500 \cm peak in the real part
of the self-energy (Fig. \ref{fig:Qazilbash}). Qazilbash \emph{et
al.} attribute this low energy feature to an electron-boson
coupling mode as found in the high-\Tc cuprates.\cite{Basov,
Hwang} Figure \ref{fig:Qazilbash}a also plots the $\omega\tau$=1
line and the scattering rate 1/$\tau(\omega)$ at 10 K lies below
this reference line, indicating well-defined quasiparticle
excitations exist in LaFePO. The overall behavior of
1/$\tau(\omega)$ is different from the optimally doped cuprates,
in which a linear frequency dependence is found.\cite{Basov} From
the inset of Fig. \ref{fig:Qazilbash}b, the mass enhancement
factor ($\omega\rightarrow$0) at T=10 K is 1.5$\pm$0.1, in good
agreement with the renormalization factor of 2 obtained from ARPES
and de Haas-van Alphen measurements.\cite{Lu, Coldea} Considering
the rather modest mass enhancement factor and its weak
\emph{T}-dependence, Qazilbash \emph{et al.} conclude LaFePO is a
moderately correlated metal.

\subsection{Ba$_{0.55}$K$_{0.45}$Fe$_2$As$_2$}\label{BaKTimusk}
Yang \emph{et al.} investigate the normal state properties for
Ba$_{0.55}$K$_{0.45}$Fe$_2$As$_2$ single crystal and also find
evidence for bosonic mode coupling.\cite{YangBaK} Since it is
difficult to separate the Drude component and the interband
transitions in K-doped BaFe$_2$As$_2$, several fits are used to
extract the intra-band contribution for the extended Drude
analysis. A bosonic excitation spectrum based on the spin
fluctuation model was used to fit the scattering rate
1/$\tau(\omega)$. The author find the coupling constant strongly
dependent on temperature ($\lambda$=0.6 when T=295 K,
$\lambda$=2.0 for T=28 K), suggesting that magnetic excitations
are responsible for the scattering. A recent ARPES study in
optimally doped Ba$_{0.6}$K$_{0.4}$Fe$_2$As$_2$ have found an
unconventional mode with an orbital selective nature: a 25 meV
mode exists for both the $\alpha$ and $\gamma$ bands, showing a
strong temperature dependence and vanishes above \Tc, while no
kink is found for the $\beta$ band.\cite{Richard} As the $\alpha$
and $\gamma$ Fermi surfaces are well connected by the SDW vector
in the parent compound, Richard \emph{et al.} therefore speculate
the kink may originate from the enhanced antiferromagnetic
fluctuations below \Tc.

\section{Phonon}
\label{phonon}

\subsection{ReFeAsO system}

\begin{figure}[t]
\begin{center}
\includegraphics[width=2.7in]{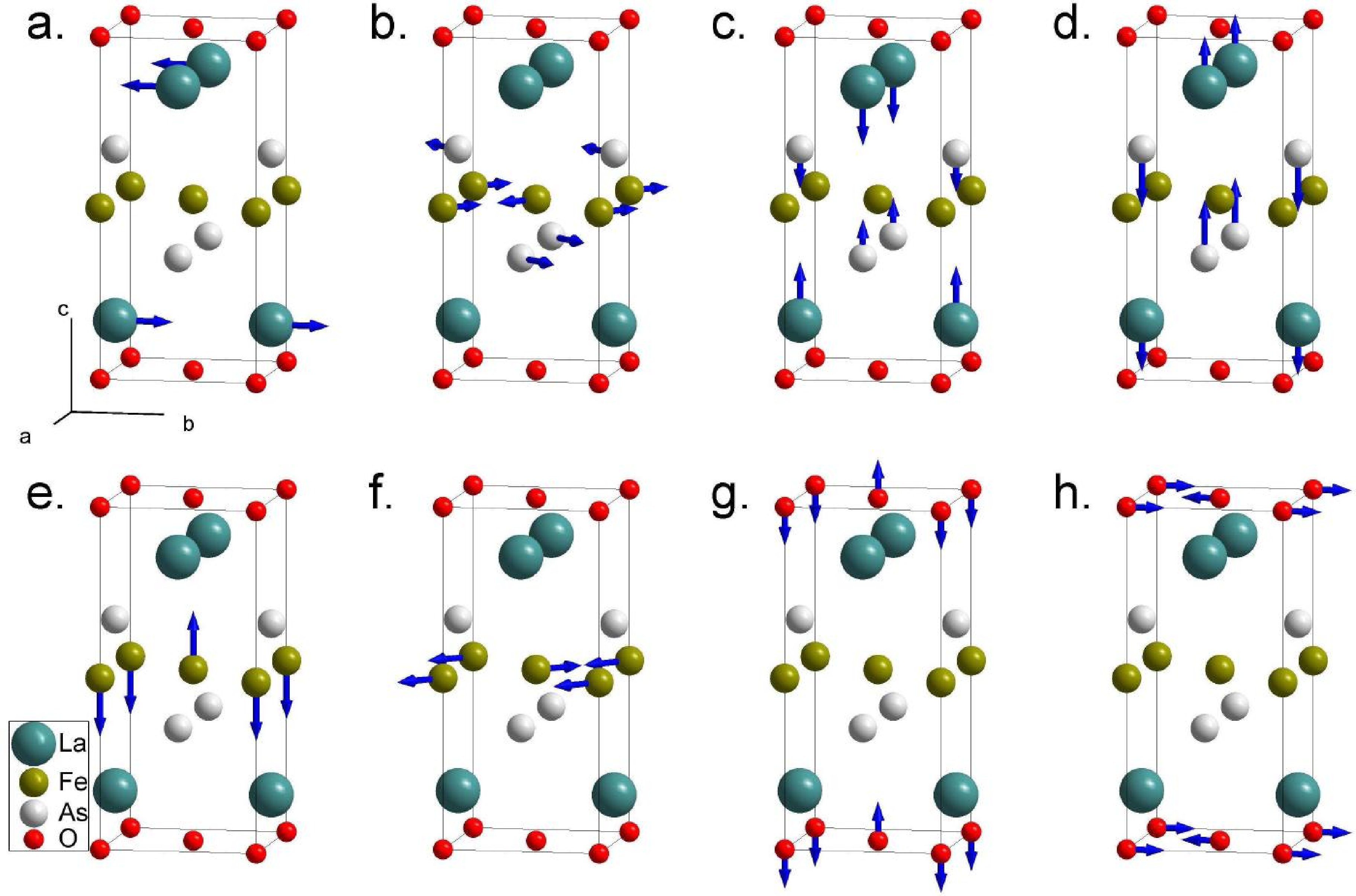}%
\caption{\label{fig:Raman1} Eight Raman active modes of LaFeAsO in
a 1111-phase representative. (a) E$_g$ of La; (b) E$_g$ of As and
Fe; (c) A$_{1g}$ of La; (d) A$_{1g}$ of As; (e) B$_{1g}$ of Fe;
(f) E$_g$ of Fe; (g) B$_{1g}$ of O; (h) E$_g$ of O.\cite{x1}
According to the Raman tensors, doubly-degenerated E$_g$ modes can
not be determined by the ab-plane Raman measurements.}
\end{center}
\end{figure}

\begin{figure}[t]
\begin{center}
\includegraphics[width=2.5in]{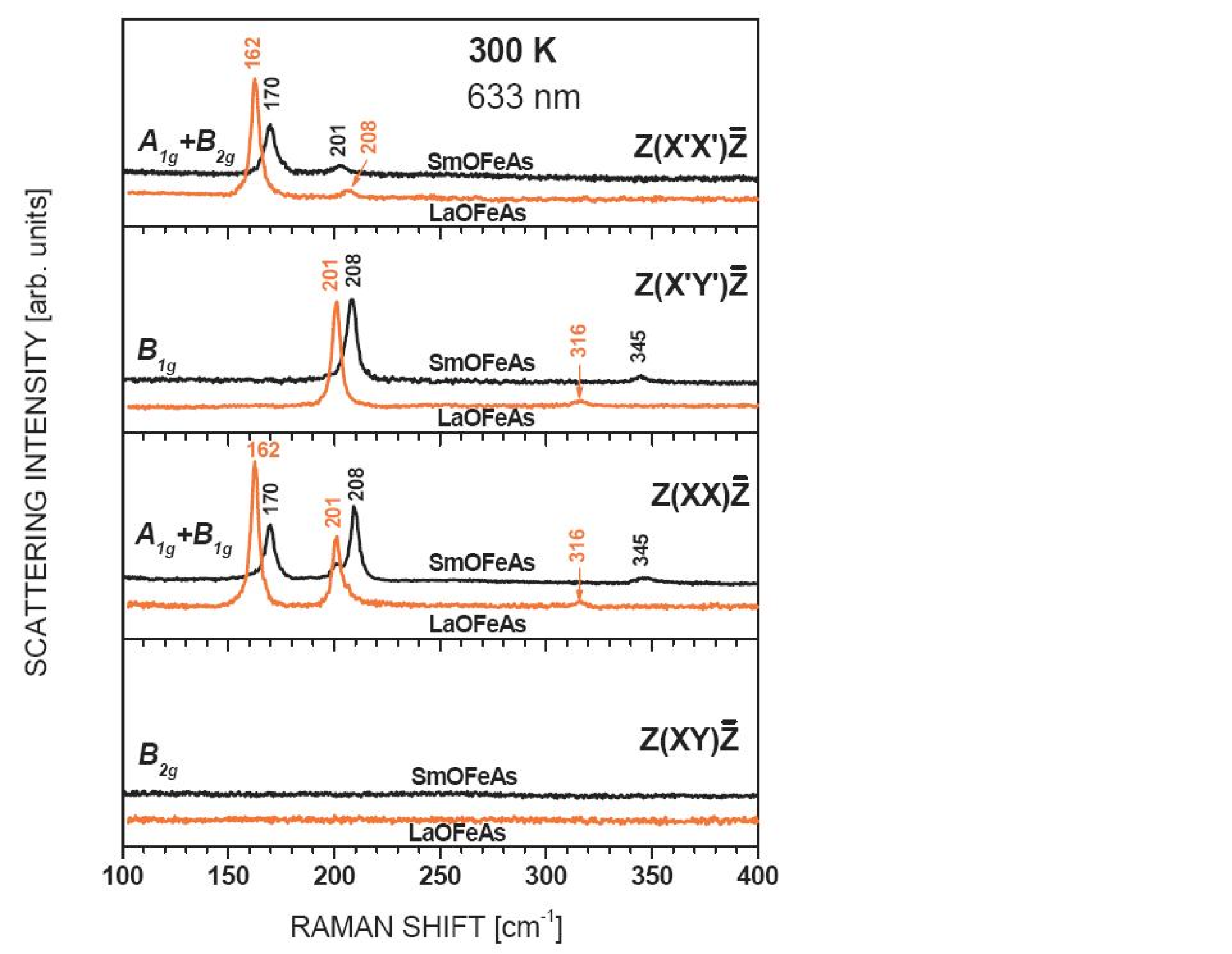}%
\vspace*{-0.20cm}%
\caption{\label{fig:Raman2} Polarized Raman spectra of LaFeAsO and
SmFeAsO. The modes located at 208/201 \cm and 170/162 \cm
corresponds to B$_{1g}$ phonons of Fe and A$_{1g}$ phonons of As
in SmFeAsO/LaFeAsO, respectively. The 345 and 316 \cm features
with weaker intensities are oxygen-related E$_g$ modes. \cite{x2}}
\end{center}
\end{figure}
The 1111-phase belongs to space group P4/nmm and point group
D$_{4h}$. The rare-earth ions, Fe, As and O(F) occupy 2c, 2b, 2c
and 2a positions, respectively. Formal group analysis gives eight
Raman active modes including two A$_{1g}$, two B$_{1g}$ and four
E$_g$ modes, and six infrared active modes including three
A$_{2u}$ and three E$_u$ modes. Representatively eight Raman
active modes of LaFeAsO are shown in Fig.
\ref{fig:Raman1}.\cite{x1} The first polarized Raman phonon
measurements on La(Sm)FeAsO were made by Hadjiev \emph{et al.}
under a microscope on very small plate-like single crystals
obtained within polycrystalline samples.\cite{x2} The four Raman
modes with vibration out of FeAs plane were identified clearly, as
shown in Fig. \ref{fig:Raman2}. Zhao \emph{et al.} reported Raman
spectra obtained in doped and undoped LaFeAsO(F) and CeFeAsO(F)
polycrystalline samples.\cite{x1} Zhang \emph{et al.} studied the
doping and temperature dependence of Raman spectra in
NdFeAsO$_{1-x}$F$_x$ polycrystalline samples.\cite{x3} Polarized
Raman spectra on NdFeAsO$_{1-x}$F$_x$ single crystals were
presented by Gallais \emph{et al.}\cite{x4} Interestingly, a
strong resonant enhancement for the Fe-related phonon modes was
found below 2 eV, which was considered to be related to the
interband transition at 2 eV as observed in optical conductivity.
Among all the Raman-active phonon modes, the B$_{1g}$ mode of iron
around 210 \cm is much more important in exploring the
superconducting and SDW mechanism for Fe-based compounds. However,
this mode is slightly sample-dependent from the above Raman
measurements.

\subsection{AFe$_2$As$_2$ system} 
\begin{figure}[t]
\begin{center}
\includegraphics[width=2.2in]{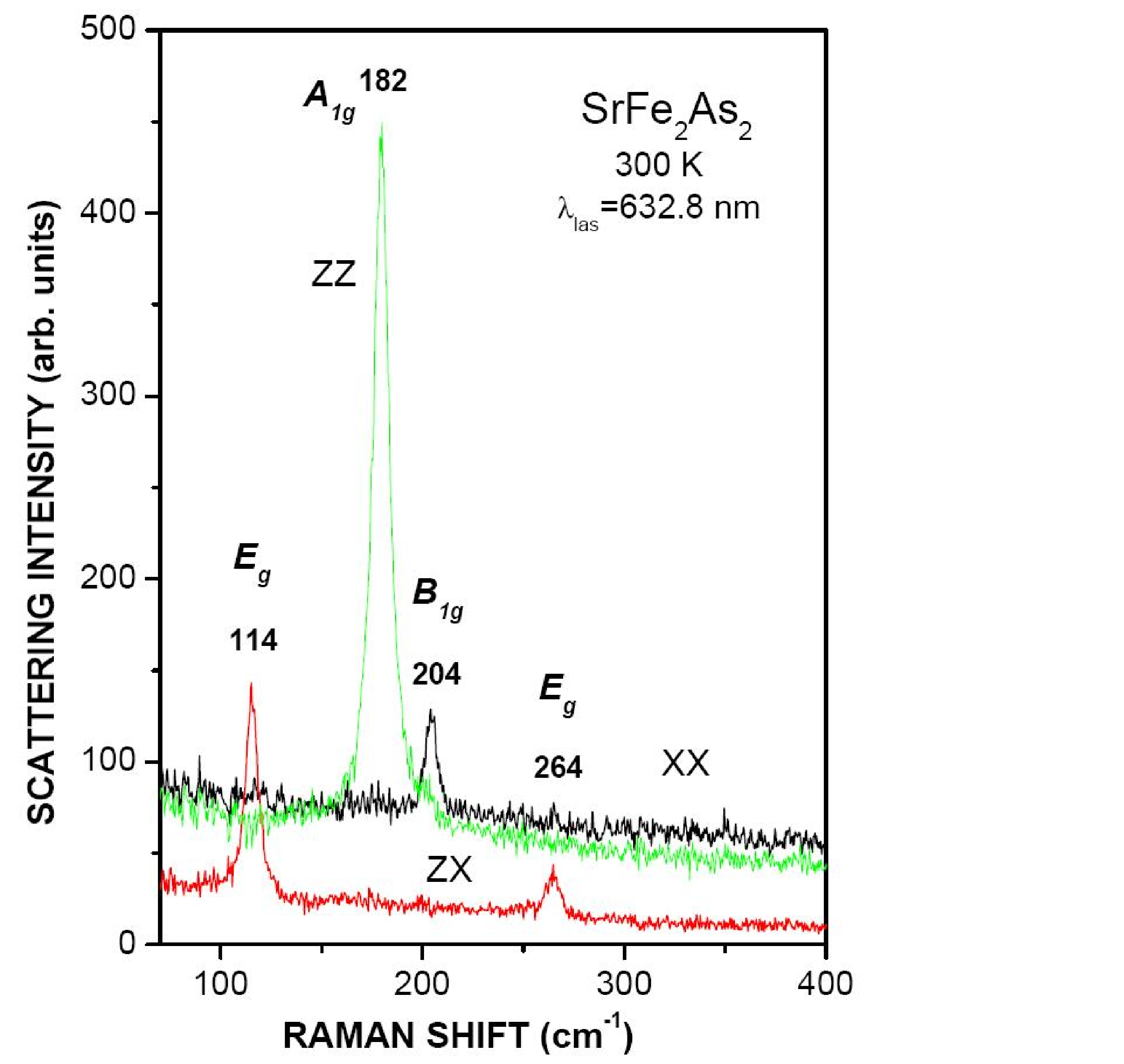}%
\vspace*{-0.20cm}%
\caption{\label{fig:Raman3} Polarized Raman spectra of
SrFe$_2$As$_2$ crystal at room temperature. The peaks at 204 and
182 \cm are one-dimensional Fe B$_{1g}$ and As A$_{1g}$ modes,
respectively. Doubly-degenerated E$_g$ modes of As and Fe were
detected on ac-surface. \cite{x5}}
\end{center}
\end{figure}

The 122-phase belongs to space
group I4/mmm and point group D$_{4h}$. Alkaline-earth ions, Fe and
As occupy 2a, 4d and 4e positions, respectively. There are four
Raman active modes including A$_{1g}$, B$_{1g}$ and two E$_g$
modes, and four infrared modes including two A$_{2u}$ and two
E$_u$ modes. Litvinchuk \emph{et al.} carried out the Raman
measurements on Sr$_{1-x}$K$_x$Fe$_2$As$_2$ crystals (shown in
Fig. \ref{fig:Raman3}),\cite{x5} and Choi et al. reported Raman
results on CaFe$_2$As$_2$ crystals.\cite{x6} Compared to the 1111
phase, B$_{1g}$ phonon frequency in the 122-phase is almost the
same as that in the 1111-phase, reflecting the fact that the
vibration of Fe is dominated by Fe-As bonds.

\subsection{Fe(SeTe) system}

\begin{figure}[t]
\begin{center}
\includegraphics[width=2.2in]{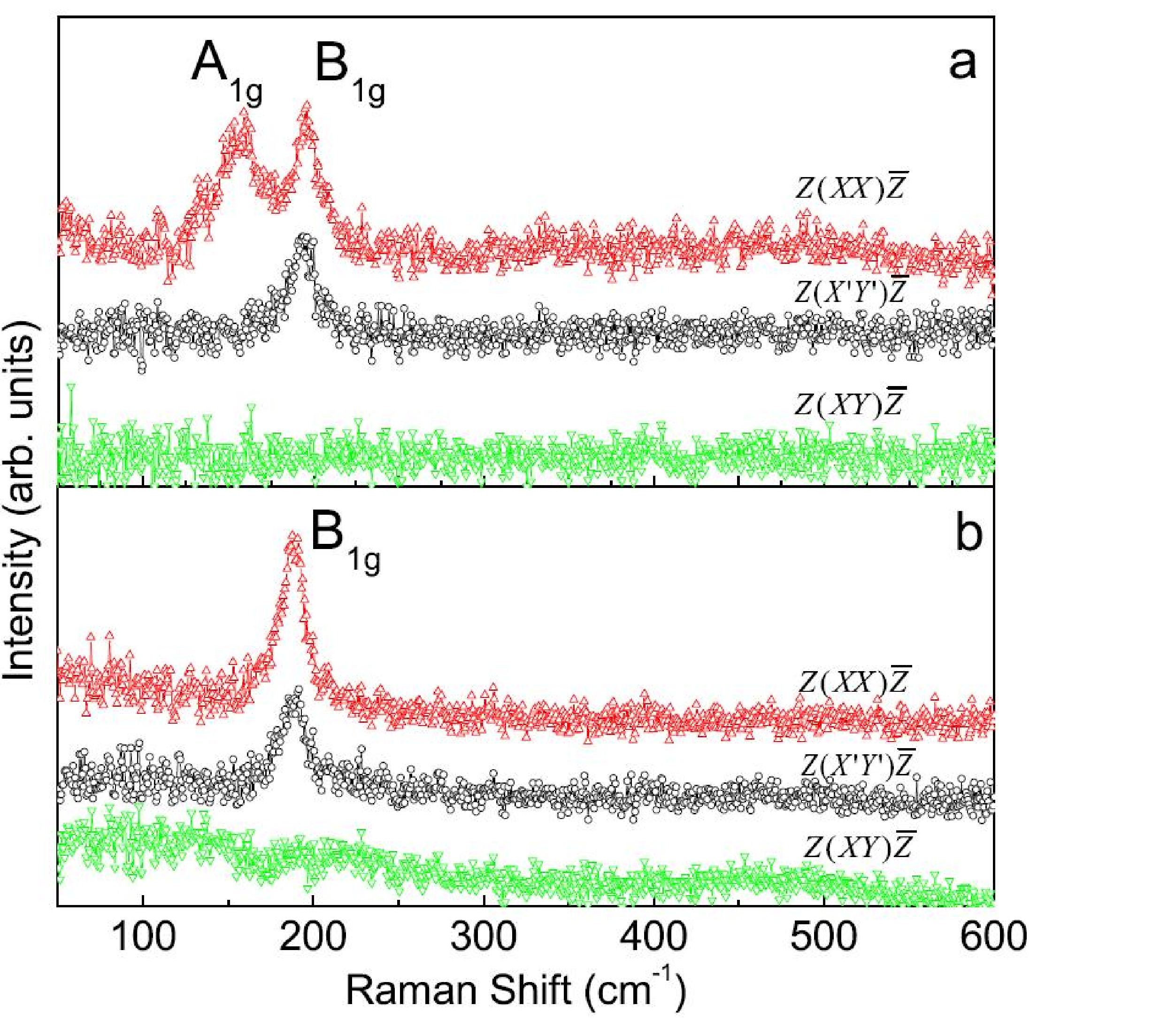}%
\vspace*{-0.20cm}%
\caption{\label{fig:Raman4} Polarized Raman spectra of
FeTe$_{0.92}$ and Fe$_{1.03}$Se$_{0.3}$Te$_{0.7}$ crystals at room
temperature. (a) Fe B$_{1g}$ and As A$_{1g}$ phonons can be found
in the parent compounds, (b) A$_{1g}$ phonons disappear in
Se-doped FeTe.\cite{x7}}
\end{center}
\end{figure}

Fe(SeTe) is a prototype of Fe-based superconductors, which has the
simplest formula without any separation layer. Its space group is
P4/nmm and point group D$_{4h}$. Fe and Se(Te) occupy 2a and 2c
positions, respectively. It has four Raman modes including
A$_{1g}$, B$_{1g}$ and two E$_g$, and two infrared modes: A$_{2u}$
and E$_u$. As shown in Fig. \ref{fig:Raman4}, A$_{1g}$ mode of Te
locates at 159 \cm. B$_{1g}$ mode of Fe is 196 \cm, slightly lower
than that of FeAs-based compounds.\cite{x7} The difference can be
explained as the change from Fe-As to Fe-Se(Te) bonds.

\subsection{Electron-phonon coupling and isotope effect}

There is no consensus on the strength of electronic correlation in
the Fe-based systems yet. The role played by electron-phonon
coupling is not clear for FeAs-based superconductors, although a
number of theoretical calculations tend to rule out the
electron-phonon coupling in the
pairing.\cite{Singh237003,Boeri,Mazin} The available Raman data do
not show any evidence of strong electron-phonon coupling in terms
of phonon lineshape and width.\cite{x1,x2,x4,x5,x6} However, a
direct isotope experiment made by Liu \emph{et al.}\cite{x8}
indicated that the isotope effect of iron on superconducting and
SDW transitions is quite large. The isotope effect is $\sim$0.4,
close to conventional BCS value. While the isotope effect of
oxygen on \Tc is negligibly small. Apparently, further experiments
are required to understand the issue. Recent X-Ray inelastic
scattering experiment may present some hints on this question. It
is suggested that a unconventional electron-phonon coupling may
exist at certain wave vectors.\cite{x9,x10,x11}

\begin{figure}[t]
\begin{center}
\includegraphics[width=3.5in]{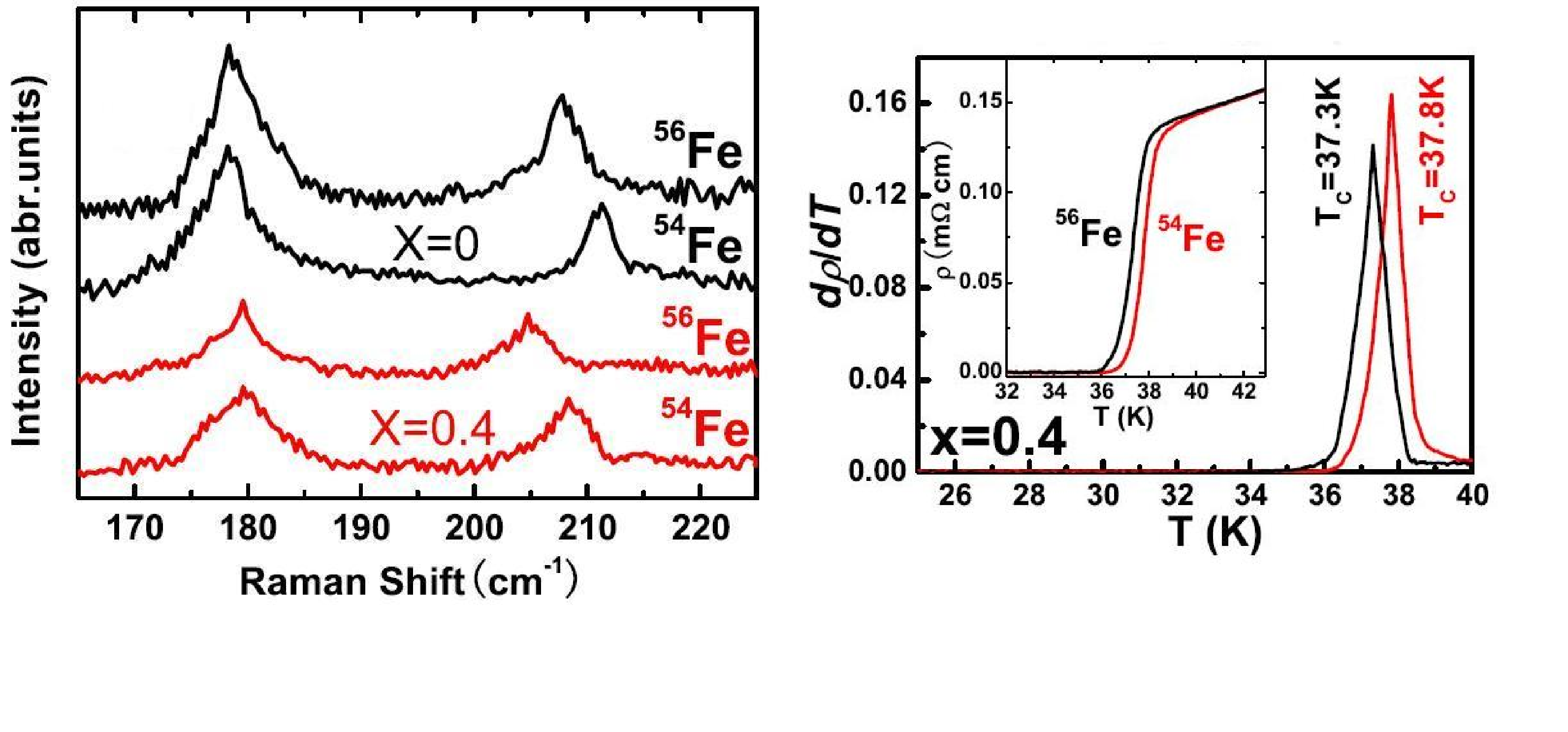}%
\vspace*{-0.20cm}%
\caption{\label{fig:Raman5} Left panel: Raman phonon shift of Fe
isotope. An obvious shift of Fe B$_{1g}$ mode can be resolved when
replacing $^{56}$Fe with $^{54}$Fe isotope. Right panel: Isotope
effect on \Tc.\cite{x8}}
\end{center}
\end{figure}

\subsection{Phonon behavior at the SDW/SC transitions}

The SDW transition in the parent compounds is usually accompanied
by a structural change. One may expect that phonon anomalies can
be observed around the SDW transition temperatures. In principle
the doubly-degenerated E$_g$ modes of Fe will be split to B$_{2g}$
and B$_{3g}$ modes after the transition from tetragonal to
orthorhombic structures. There is no such splitting reported yet.
The estimated splitting could be too small to be determined.
However Raman measurements on SrFe$_2$As$_2$ crystals do not show
such anomalies for the in-plane modes.\cite{x5} On the other hand,
Raman measurements on CaFe$_2$As$_2$ crystals indicate that a
large jump of $\sim$ 4 \cm of Fe B$_{1g}$ phonon occurs around the
structural transition temperature 173 K.\cite{x6} The jump can not
be explained by lattice distortion. It is suggested that the jump
could be related to the change of electronic density at Fermi
surface modified by structural transitions. For superconducting
Sr$_{1-x}$K$_x$Fe$_2$As$_2$ and NdFeAsO$_{1-x}$F$_x$ crystals, no
phonon anomalies can be seen cross \Tc.\cite{x4,x5}

As shown in Fig. \ref{fig:LaFeAsOEPL}, five pronounced phonon
modes at 100, 247, 265, 336, 438 \cm dominate the far-infrared
reflectivity R($\omega$) for polycrystalline
LaFeAsO.\cite{DongEPL} According to calculated PDOS (phonon
density of states), the phonon modes above 300 \cm are strongly
oxygen-derived, while the phonons below 300 \cm are a mixture of
La, Fe, and As vibration modes.\cite{Singh237003, Boeri} All
phonon modes seems unchanged across the 150 K transition, which is
later identified as a structural transition (from P4/nmm to Cmma)
by neutron scattering.\cite{Clarina} Yildirim calculates the
IR-active phonon modes and finds such a structural distortion will
not introduce new phonons at the zone center $\Gamma$, but just
split the doubly degenerate modes into nondependence ones and the
splitting is smaller than 0.2 meV, hence no new phonons appear in
the optical reflectivity for LaFeAsO after the structural
transition.\cite{Yildirim}

\begin{figure}[t]
\begin{center}
\includegraphics[width=2.7in]{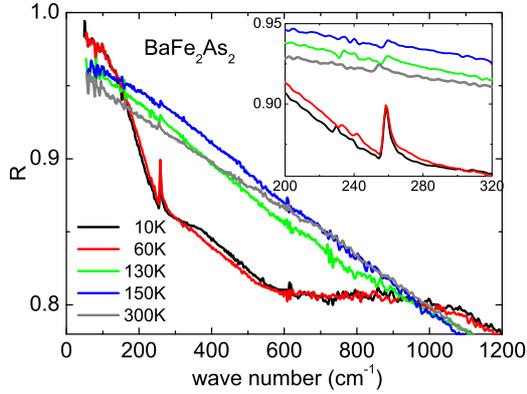}%
\vspace*{-0.20cm}%
\caption{\label{fig:phonon} Reflectivity R($\omega$) for
BaFe$_2$As$_2$ single crystal below 1200 \cm. The inset is an
enlarged plot to clearly illustrate the infrared-active phonon
around 259 \cm. This phonon exists for all temperatures.}
\end{center}
\end{figure}

Figure \ref{fig:phonon} is the ab-plane optical reflectivity for
single crystalline BaFe$_2$As$_2$. Only one infrared-active phonon
mode around 259 \cm is found. Similar to the ReFeAsO case, here no
new phonons appear in the spin-density wave state. As shown in the
inset figure, this 259 \cm phonon exists even in 300 K, which
apparently turns sharper in 10 K. This phonon can also be found in
optimally doped (Ba,K)Fe$_2$As$_2$ (see Fig. \ref{fig:BaKR}) but
is much weaker due to a better screening of itinerant carriers
after K doping.

\section{Conclusion}
\label{conclusion}

Almost one year has passed since the discovery of a 26 K
superconductor LaFeAsO$_{1-x}$F$_x$. Besides the exploration for
new types of Fe-based superconductor with higher \Tc, great
efforts have been devoted to finding out how and why
superconductivity emergences from Fe 3\emph{d} electrons. In
retrospect, the most convinced and important spectroscopic results
were obtained from 1111-type polycrystals and 122-type single
crystals for the in-plane optical response. From the SDW gap in
the metallic parent compounds, and the superconducting gap for the
doped systems, we know the Fe-based superconductors (especially
the 122 system) are different from the high-\Tc cuprates but more
or less closer to the standard BCS superconductors. However,
information on the charge dynamics for the c-axis, or the symmetry
of the superconducting gap in single crystalline 1111-type
Fe-based superconductor is still lacking. Moreover, investigations
on the electronic correlation and the pesudogap-like feature in
Fe-based compounds requires further studies, especially
spectroscopic investigations on high-quality single crystals.
Anyhow, some basic conclusions on Fe-based superconductors and
their parent compounds can be made based on current optical and
Raman data, which are briefly summarized as below.

\emph{Itinerant system.} Almost all Fe-based superconducting
systems are metallic in nature. In particular, the parent compound
is not an antiferromagnetic Mott insulator, but a semimetal with
an SDW instability. The plasma frequency in the normal phase
(above T$_{SDW}$) is about 1.5 eV. When the nesting condition is
weakened by either carrier doping or application of pressure,
superconductivity wins in the ground state.

\emph{Multi-band effect.} Fe-based superconductors are multi-band
systems, which are different from the situation in cuprates
(one-band). The BCS double-gap is observed in
Ba$_{0.6}$K$_{0.4}$Fe$_2$As$_2$ from ARPES measurement. With a
detailed analysis of the optical conductivity, similar two-gap
superconductivity can also be found in optical response. Another
optical evidence for the multi-band property in iron arsenide is
the SDW double-gap in AFe$_2$As$_2$ (Sr, Ba) parent compounds.

\emph{s-wave superconductivity.} From optical measurement on
polycrystalline ReFeAsO$_{1-x}$F$_x$, signature of the
superconducting gap is observed at the far-infrared as an up-turn
in reflectivity below \Tc. More convinced evidence of the gap
symmetry comes from optical data on single crystal. For optimally
doped (Ba,K)Fe$_2$As$_2$, an s-wave pairing lineshape is found for
the superconducting gap, thus different from the \emph{d}-wave gap
in cuprates.

\emph{Electron-phonon coupling} The large isotope effect implies
that electron-phonon coupling may be strong in the system.
However, considering its high T$_c$, it is unlikely that
electron-phonon coupling dominates the pairing as that in
conventional BCS superconductors. This issue needs further
experimental and theoretical efforts in the future.

\section{Acknowledgement}
This work is supported by the National Science Foundation of
China, the Knowledge Innovation Project of the Chinese Academy of
Sciences, and the 973 project of the Ministry of Science and
Technology of China.


\begin{thebibliography}{99}

\bibitem{Kamihara08} Y. Kamihara, T. Watanabe, M. Hirano, and
H. Hosono, J. Am. Chem. Soc. 130, 3296 (2008).

\bibitem{XHChen} X. H. Chen, T. Wu, G. Wu, R. H. Liu, H.
Chen, and D. F. Fang, Nature \textbf{453}, 761 (2008).

\bibitem{ChenPRLCe} G. F. Chen, Z. Li, D. Wu, G. Li, W. Z. Hu, J. Dong, P. Zheng, J. L. Luo, and N. L.
Wang, Phys. Rev. Lett. \textbf{100}, 247002 (2008).

\bibitem{Ren1} Z. A. Ren, W. Lu, J. Yang, W. Yi, X. L. Shen,
Z. C. Li, G. C. Che, X. L. Dong, L. L. Sun, F. Zhou, and Z. X.
Zhao, Chin. Phys. Lett. \textbf{25}, 2215 (2008)

\bibitem{DongEPL} J. Dong, H. J. Zhang, G. Xu, Z. Li, G. Li, W. Z. Hu, D. Wu,
G. F. Chen, X. Dai, J. L. Luo, Z. Fang, and N. L. Wang, Europhys.
Lett. \textbf{83}, 27006 (2008).

\bibitem{Clarina} Clarina de la Cruz, Q. Huang, J. W. Lynn, Jiying Li, W. Ratcliff II, J. L. Zarestky, H. A. Mook, G. F. Chen,
J. L. Luo, N. L. Wang, and Pengcheng Dai, Nature \textbf{453}, 899
(2008).

\bibitem{Rotter1} Marianne Rotter, Marcus Tegel, Dirk Johrendt, Inga Schellenberg, Wilfried Hermes, and Rainer P\"{o}ttgen, Phys. Rev. B \textbf{78}, 020503(R) (2008).

\bibitem{Rotter2} Marianne Rotter, Marcus Tegel, and Dirk Johrendt, Phys. Rev. Lett. \textbf{101}, 107006
(2008).

\bibitem{Krellner}  C. Krellner, N. Caroca-Canales, A. Jesche, H. Rosner, A. Ormeci, and C.
Geibel, Phys. Rev. B \textbf{78}, 100504(R) (2008) .

\bibitem{Chen2} G. F. Chen, Z. Li, G. Li, W. Z. Hu, J. Dong, P. Zheng, N. L. Wang and J. L. Luo, Chin. Phys. Lett. \textbf{25}, 3403 (2008).

\bibitem{Huang} Q. Huang, Y. Qiu, Wei Bao, J.W. Lynn, M.A. Green, Y. Chen, T. Wu,
G. Wu, and X.H. Chen, Phys. Rev. Lett. \textbf{101}, 257003
(2008).

\bibitem{Zhao} Jun Zhao, W. Ratcliff II, J. W. Lynn, G. F. Chen, J. L. Luo, N. L.
Wang, Jiangping Hu, and Pengcheng Dai, Phys. Rev. B \textbf{78},
140504(R) (2008).

\bibitem{Ni} N. Ni, S. L. Bud'ko, A. Kreyssig, S. Nandi, G. E. Rustan, A. I. Goldman, S. Gupta, J. D. Corbett, A. Kracher, and P. C.
 Canfield, Phys. Rev. B \textbf{78}, 014507 (2008).

\bibitem{Chen3} G. F. Chen, Z. Li, J. Dong, G. Li, W. Z. Hu, X. D. Zhang, X. H.
Song, P. Zheng, N. L. Wang, and J. L. Luo, Phys. Rev. B
\textbf{78}, 224512 (2008).

\bibitem{Jeevan} H. S. Jeevan, Z. Hossain, Deepa Kasinathan, H. Rosner, C. Geibel, and P.
Gegenwart, Phys. Rev. B \textbf{78}, 052502 (2008)

\bibitem{ZhiRen} Zhi Ren, Zengwei Zhu, Shuai Jiang, Xiangfan Xu, Qian Tao, Cao Wang, Chunmu Feng, Guanghan Cao, and
Zhu'an Xu, Phys. Rev. B \textbf{78}, 052501 (2008).

\bibitem{FeSe} Fong-Chi Hsu, Jiu-Yong Luo, Kuo-Wei Yeh, Ta-Kun Chen, Tzu-Wen Huang, Phillip M. Wu, Yong-Chi Lee, Yi-Lin Huang, Yan-Yi Chu, Der-Chung Yan, and Maw-Kuen
Wu, Proc. Nat. Acad. Sci. \textbf{105}, 14262(2008).

\bibitem{FeSeTe} Kuo-Wei Yeh, Tzu-Wen Huang, Yi-Lin Huang, Ta-Kun Chen, Fong-Chi Hsu, Phillip M. Wu, Yong-Chi Lee, Yan-Yi Chu, Chi-Lian Chen, Jiu-Yong Luo, Der-Chung Yan, and Maw-Kuen
Wu, Europhys. Lett. \textbf{84}, 37002(2008).

\bibitem{CQJin} X.C. Wang, Q.Q. Liu, Y.X. Lv, W.B. Gao, L.X. Yang, R.C. Yu,
F.Y. Li, and C.Q. Jin, Solid Sate Communications, \textbf{148},
538 (2008).

\bibitem{Tapp} J.H. Tapp, Z. Tang, B. Lv, K. Sasmal, B. Lorenz, C.W. Chu,
and A.M. Guloy, Phys. Rev. B \textbf{78}, 060505(R) (2008).

\bibitem{Pitcher} Michael J. Pitcher, Dinah R. Parker, Paul Adamson, Sebastian J. C. Herkelrath, Andrew T. Boothroyd, and Simon J.
 Clarke, Chemical Communications 5918 - 5920 (2008).

\bibitem{Sadovskii} M. V. Sadovskii, arXiv:0812.0302.

\bibitem{ChenPRLLa} G. F. Chen, Z. Li, G. Li, J. Zhou, D. Wu, J. Dong, W. Z. Hu, P. Zheng, Z. J. Chen, H. Q. Yuan, J. Singleton, J. L. Luo, and N. L.
Wang, Phys. Rev. Lett. \textbf{101}, 057007 (2008).

\bibitem{Dubroka} A. Dubroka, K.W. Kim, M. R\"{o}ssle, V. K. Malik, A. J. Drew, R. H. Liu, G. Wu, X. H. Chen, and C.
Bernhard, Phys. Rev. Lett. \textbf{101}, 097011 (2008).

\bibitem{Hu122} W. Z. Hu, J. Dong, G. Li, Z. Li, P. Zheng, G. F. Chen, J. L. Luo, and N. L.
Wang, Phys. Rev. Lett. \textbf{101}, 257005(2008).

\bibitem{BaKPRL} G. Li, W. Z. Hu, J. Dong, Z. Li, P. Zheng, G. F.
Chen, J. L. Luo, and N. L. Wang, Phys. Rev. Lett. \textbf{101},
107004 (2008).

\bibitem{Yildirim} T. Yildirim, Phys. Rev. Lett. \textbf{101}, 057010 (2008).

\bibitem{Ma} Fengjie Ma, Zhong-Yi Lu, and Tao Xiang, Phys. Rev. B \textbf{78}, 224517 (2008).

\bibitem{Fang} Chen Fang, Hong Yao, Wei-Feng Tsai, JiangPing Hu, and Steven A.
Kivelson, Phys. Rev. B \textbf{77}, 224509 (2008).

\bibitem{Xu} Cenke Xu, Markus Mueller, and Subir Sachdev, Phys. Rev. B \textbf{78}, 020501(R) (2008).

\bibitem{Marini} C. Marini, C. Mirri, G. Profeta, S. Lupi, D. Di Castro, R. Sopracase, P. Postorino, P. Calvani, A. Perucchi, S. Massidda, G. M. Tropeano, M. Putti, A. Martinelli, A. Palenzona, and P.
Dore, arXiv:0810.2176.

\bibitem{Tropeano}  M. Tropeano, C. Fanciulli, C. Ferdeghini, D. Marr\`{e}, A.S. Siri, M. Putti, A. Martinelli, M. Ferretti, A. Palenzona, M. R. Cimberle, C. Mirri, S. Lupi, R. Sopracase, P. Calvani, and A.
Perucchi, arXiv:0809.3500.

\bibitem{ZhaoNatMat}  J. Zhao, Q. Huang, C. de la Cruz, S. Li, J. W. Lynn, Y. Chen, M. A. Green, G. F. Chen,
G. Li, Z. Li, J. L. Luo, N. L. Wang, and Pengcheng Dai, Nature
Materials \textbf{7}, 953 (2008).


\bibitem{Singh237003}D. J. Singh and M.-H. Du, Phys. Rev. Lett.
\textbf{100}, 237003(2008).

\bibitem{Boeri} L. Boeri, O. V. Dolgov, and A. A. Golubov, Phy.
Rev. Lett. \textbf{101}, 026403(2008).

\bibitem{BorisLa} A. V. Boris, N. N. Kovaleva, S. S. A. Seo, J. S. Kim,
P. Popovich, Y. Matiks, R. K. Kremer, and B. Keimer, Phys. Rev.
Lett. \textbf{102,} 027001 (2009).

\bibitem{HuUnpub} W. Z. Hu, et al. unpublished data.

\bibitem{SinghBa122} D. J. Singh, Phys. Rev. B \textbf{78}, 094511
(2008).

\bibitem{Feng} L. X. Yang, Y. Zhang, H. W. Ou, J. F. Zhao, D. W. Shen, B. Zhou,
J. Wei, F. Chen, M. Xu, C. He, Y. Chen, Z. D. Wang, X. F. Wang, T.
Wu, G. Wu, X. H. Chen, M. Arita, K. Shimada, M. Taniguchi, Z. Y.
Lu, T. Xiang, and D. L. Feng, arXiv:0806.2627.

\bibitem{Kaminski} C. Liu, G. D. Samolyuk, Y. Lee, N. Ni, T. Kondo, A. F.
Santander-Syro, S. L. Bud'ko, J. L. McChesney, E. Rotenberg, T.
Valla, A. V. Fedorov, P. C. Canfield, B. N. Harmon, and A.
Kaminski, Phys. Rev. Lett. \textbf{101}, 177005 (2008).

\bibitem{Hsieh} D. Hsieh, Y. Xia, L. Wray, D. Qian, K.K. Gomes, A.
Yazdani, G.F. Chen, J.L. Luo, N.L. Wang, and M.Z. Hasan,
arXiv:0812.2289.

\bibitem{Dressel} M. Dressel, and G. Gr\"{u}ner, \emph{Electrodynamics of Solids}, Cambridge University Press (2002).

\bibitem{Sebastian} Suchitra E. Sebastian, J. Gillett, N. Harrison, C. H. Mielke, S.
K. Goh, P. H. C. Lau, and G. G. Lonzarich, J. Phys.: Condens.
Matter \textbf{20}, 422203 (2008).

\bibitem{Si} Q. Si, and E. Abrahams, Phys. Rev. Lett. \textbf{101}, 076401 (2008).

\bibitem{Wu} J. Wu, P. Phillips, and A. H. Castro Neto,
Phys. Rev. Lett. \textbf{101}, 126401 (2008).

\bibitem{Mazin}I. I. Mazin, D. J. Singh, M. D. Johannes, and M. H.
Du, Phys. Rev. Lett. \textbf{101}, 057003 (2008).

\bibitem{Cvetkovic} Vladimir Cvetkovic and Zlatko Tesanovic,
arXiv:0804.4678.

\bibitem{HYLiu} Haiyun Liu, Wentao Zhang, Lin Zhao, Xiaowen Jia, Jianqiao Meng, Guodong Liu, Xiaoli Dong, G. F. Chen, J. L. Luo, N. L. Wang, Wei Lu, Guiling Wang, Yong Zhou, Yong Zhu, Xiaoyang Wang, Zuyan Xu, Chuangtian Chen, and X. J.
Zhou, Phy. Rev. B \textbf{78}, 184514 (2008)

\bibitem{DingBaK} H. Ding, K. Nakayama, P. Richard, S. Souma, T. Sato, T. Takahashi, M. Neupane, Y.-M. Xu, Z.-H. Pan, A.V. Federov, Z. Wang, X. Dai, Z. Fang, G.F. Chen, J.L. Luo, and N.L.
Wang, arXiv:0812.0534.

\bibitem{Sato} T. Sato, K. Nakayama, Y. Sekiba, P. Richard, Y.-M. Xu, S. Souma,
T. Takahashi, G. F. Chen, J. L. Luo, N. L. Wang, and H. Ding,
arXiv:0810.3047.

\bibitem{Sekiba} Y. Sekiba, T. Sato, K. Nakayama, K. Terashima, P. Richard, J. H. Bowen, H. Ding, Y.-M. Xu, L. J. Li, G. H. Cao, Z.-A. Xu, and T.
Takahashi1, arXiv:0812.4111.

\bibitem{Terashima} K. Terashima, Y. Sekiba, J. H. Bowen, K. Nakayama, T. Kawahara, T. Sato, P. Richard, Y.-M. Xu, L. J. Li, G. H. Cao, Z.-A. Xu, H. Ding, and T.
Takahashi, arXiv:0812.3704.

\bibitem{Bao} W. Bao, Y. Qiu, Q. Huang, A. Green, P. Zajdel,
M.R. Fitzsimmons, M. Zhernenkov, M. Fang, B. Qian, E.K. Vehstedt,
J. Yang, H.M. Pham, L. Spinu, and Z.Q. Mao, arXiv:0809.2058.

\bibitem{Li} Shiliang Li, Clarina de la Cruz, Q. Huang, Y. Chen, J. W. Lynn,
Jiangping Hu, Yi-Lin Huang, Fong-chi Hsu, Kuo-Wei Yeh, Maw-Kuen
Wu, and Pengcheng Dai, arXiv:0811.0195.

\bibitem{Wu01} F.C. Hsu, T.Y. Luo, K.W. Yeh, T.K. Chen, T.W. Huang, Phillip M. Wu,
Y.C.  Lee, Y.L. Huang, Y.Y. Chu, D.C. Yan, and M.K. Wu, Proc.
Natl. Acad. Sci. USA. \textbf{105}, 14262 (2008).

\bibitem{Mao} M.H. Fang, H.M. Pham, B. Qian, T.J. Liu, E.K. Vehstedt,
Y. Liu, L. Spinu, and Z.Q. Mao, Phys. Rev. B \textbf{78}, 224503
(2008).

\bibitem{Wu02} K.W. Yeh, T.W. Huang, Y.L. Huang,
T.K. Chen, F.C. Hsu, Phillip M. Wu, Y.C. Lee, Y.Y Chu, C.L. Chen,
J.Y. Luo, D.C. Yan, and M.K. Wu, arXiv:0808.0474

\bibitem{Mizuguchi} Y. Mizuguchi, F. Tomioka, S. Tsuda, T. Yamaguchi, and Y. Takano,
Appl. Phys. Lett. \textbf{93}, 152505 (2008).

\bibitem{ChenFeTe} G. F. Chen, Z. G. Chen, J. Dong, W. Z. Hu, G. Li, X. D. Zhang, P. Zheng, J. L. Luo, and N. L.
Wang, arXiv:0811.1489.

\bibitem{XiaFeTe} Y. Xia, D. Qian, L. Wray, D. Hsieh, G.F. Chen, J.L. Luo, N.L. Wang, and M.Z.
 Hasan, arXiv:0901.1299.

\bibitem{JZhang} Lijun Zhang, D.J. Singh, and Mao-Hua Du, arXiv:0810.3274.

\bibitem{Mirzaei} S.I. Mirzaei, V. Guritanu, A. B. Kuzmenko, C. Senatore, D. van
der Marel, G. Wu, R. H. Liu, and X. H. Chen, arXiv:0806.2303.

\bibitem{DrechslerLa} S.-L. Drechsler, M. Grobosch, K. Koepernik, G. Behr, A. K\"{o}hler,
J. Werner, A. Kondrat, N. Leps, Ch. Hess, R. Klingeler, R. Schuster, B. B\"{u}chner, and M.
Knupfer, Phys. Rev. Lett. \textbf{101}, 257004(2008).


\bibitem{Tinkham} M. Tinkham, \emph{Introduction to Superconductivity}(2nd Ed.), (McGraw-Hill, New
York, 1996).

\bibitem{Homes05} C. C. Homes, S. V. Dordevic, T. Valla, and M.
Strongin, Phys. Rev. B \textbf{72}, 134517 (2005).

\bibitem{Homes04} C. C. Homes, S. V. Dordevic, D. A. Bonn, Ruixing Liang, and W. N. Hardy, Phys. Rev. B \textbf{69}, 024514 (2004).

\bibitem{Ferrell58} R. A. Ferrell and R. E. Glover,III, Phys. Rev.
\textbf{109}, 1398 (1958).

\bibitem{Tinkham59} M. Tinkham and R. A. Ferrell, Phys. Rev. Lett.
\textbf{2}, 331 (1959).

\bibitem{Homes06} C. C. Homes, R. P. S. M. Lobo, P. Fournier, A.
Zimmers, and R. L. Greene, Phys. Rev. B \textbf{74}, 214515
(2006).

\bibitem{YangBaK} J. Yang, D. H\"{u}vonen, U. Nagel, T. R\~{o}\~{o}m, N. Ni, P. C. Canfield, S. L. Bud'ko, J.P. Carbotte, and T.
Timusk, arXiv:0807.1040.

\bibitem{Molegraaf} H. J. A. Molegraaf, C. Presura, D. van der Marel, P. H.
Kes, and M. Li, Science \textbf{295}, 2239 (2002).

\bibitem{Bontemps} F. Santander-Syro, R. P. S. M. Lobo, N. Bontemps, Z. Konstantinovic, Z. Z. Li,
and H. Raffy, Europhys. Lett. \textbf{62}, 568 (2003).

\bibitem{DingARPES}  H. Ding, P. Richard, K. Nakayama, T. Sugawara, T. Arakane, Y. Sekiba, A. Takayama, S. Souma, T. Sato, T. Takahashi, Z. Wang, X. Dai, Z. Fang, G. F. Chen, J. L. Luo, and N. L.
Wang, Europhys. Lett. \textbf{83}, 47001 (2008).

\bibitem{ZhaoARPES}  Lin Zhao, Haiyun Liu, Wentao Zhang, Jianqiao Meng, Xiaowen Jia, Guodong Liu, Xiaoli Dong, G. F. Chen, J. L. Luo, N. L. Wang, Guiling Wang, Yong Zhou, Yong Zhu, Xiaoyang Wang, Zhongxian Zhao, Zuyan Xu, Chuangtian Chen, and X. J.
Zhou, Chin. Phys. Lett. \textbf{25}, 4402(2008)

\bibitem{Nakayama} K. Nakayama, T. Sato, P. Richard, Y.-M. Xu, Y. Sekiba,
S. Souma, G. F. Chen, J. L. Luo, N. L. Wang, H. Ding, and T.
Takahashi, arXiv:0812.0663.

\bibitem{Gruner} G. Gr\"{u}ner, \emph{Density Waves in Solids} (Addison-Weslsy, Reading, MA,
1994).

\bibitem{Stroud} Sergey V. Barabash, and David Stroud, Phys. Rev. B \textbf{66}, 172501 (2002).

\bibitem{Qazilbash} M. M. Qazilbash, J. J. Hamlin, R. E. Baumbach, M. B. Maple, and D. N.
Basov, arXiv:0808.3748.

\bibitem{LaFePO} Y. Kamihara, H. Hiramatsu, M. Hirano, R.
Kawamura, H. Yanagi, T. Kamiya, and H. Hosono, J. Am. Chem. Soc.
\textbf{128}, 10012 (2006).

\bibitem{LaFePOsingle}J. J. Hamlin, R. E. Baumbach, D. A. Zocco, T. A. Sayles, and M. B.
Maple, J. Phys.: Condens. Matter \textbf{20}, 365220 (2008).

\bibitem{Baumbach} R. E. Baumbach, J. J. Hamlin, L. Shu, D. A. Zocco,
N. M. Crisosto, and M. B. Maple, arXiv:0812.0774.

\bibitem{Fletcher} J.D. Fletcher, A. Serafin, L. Malone, J. Analytis, J-H Chu, A.S. Erickson, I.R. Fisher, and A.
Carrington, arXiv:0812.3858.

\bibitem{Lu} D. H. Lu, M. Yi, S.-K. Mo, A. S. Erickson, J. Analytis, J.-H. Chu,
D. J. Singh, Z. Hussain, T. H. Geballe, I. R. Fisher, and Z.-X.
Shen, Nature \textbf{455}, 81(2008).

\bibitem{Basov} D. N. Basov and T. Timusk, Rev. Mod. Phys.
\textbf{77}, 721 (2005).

\bibitem{Hwang} J. Hwang, T. Timusk, and G. D. Gu, Nature
\textbf{427}, 714 (2004).

\bibitem{Coldea} A.I. Coldea, J.D. Fletcher, A. Carrington,
J.G. Analytis, A.F. Bangura, J.-H. Chu, A. S. Erickson, I.R.
Fisher, N.E. Hussey, and R.D. McDonald, arXiv:0807.4890.

\bibitem{Richard} P. Richard, T. Sato, K. Nakayama, S. Souma, T. Takahashi, Y.-M.
Xu, G. F. Chen, J. L. Luo, N. L. Wang, and H. Ding, Phys. Rev.
Lett. \textbf{102}, 047003 (2009).

\bibitem{x1} S. C. Zhao, D. Hou, Y. Wu, T. L. Xia, A. M. Zhang, G. F.
Chen, J. L. Luo, N. L. Wang, J. H. Wei, Z. Y. Lu, and Q. M. Zhang,
Supercond. Sci. Technol. 22, 015017(2009).

\bibitem{x2} V. G. Hadjiev, M. N. Iliev, K. Sasmal, Y. -Y. Sun, and C. W.
Chu, Phys. Rev. B 77,  220505 (2008).

\bibitem{x3} L. Zhang, T. Fujita, F. Chen, D. L. Feng, S. Maekawa, and M.
W. Chen, arXiv:0809.1474.

\bibitem{x4} Y. Gallais, A. Sacuto, M. Cazayous, P. Cheng, L. Fang, and H.
H. Wen, Phys. Rev. B 78 (2008) 132509.

\bibitem{x5} A. P. Litvinchuk, V. G. Hadjiev, M. N. Iliev, Bing Lv, A. M.
Guloy, and C. W. Chu, Phys. Rev. B 78, 060503(2008).

\bibitem{x6} K.-Y. Choi, D. Wulferding, P. Lemmens, N. Ni, S. L. Bud'ko
and P. C. Canfield, arXiv:0810.2040.

\bibitem{x7} Tian-Long Xia, D. Hou, S. C. Zhao, A. M. Zhang, G. F. Chen,
J. L. Luo, N. L. Wang, J. H. Wei, Z. -Y. Lu, and Q. M. Zhang,
arXiv:0811.2350.

\bibitem{x8}  Tatsuo Fukuda, Alfred Q. R. Baron, Shin-ichi Shamoto, Motoyuki Ishikado, Hiroki Nakamura, Masahiko Machida, Hiroshi Uchiyama, Satoshi Tsutsui, Akira Iyo, Hijiri Kito, Jun'ichiro Mizuki, Masatoshi Arai, Hiroshi Eisaki, and Hideo Hosono, J. Phys. Soc.
Jpn. \textbf{77} 103715 (2008).

\bibitem{x9} M. Le Tacon, M. Krisch, A. Bosak, J.-W. G. Bos, and S.
Margadonna, Phys. Rev. B 78, 140505(R) (2008).

\bibitem{x10} D. Reznik, K. Lokshin, D. C. Mitchell, D. Parshall, W.
Dmowski, D. Lamago, R. Heid, K.-P- Bohnen, A. S. Sefat, M. A.
McGuire, B. C. Sales, D. G. Mandrus, A. Asubedi, D. J. Singh, A.
Alatas, M. H. Upton, A. H. Said, Yu. Shvyd'ko, and T. Egami,
arXiv: 0810.4941.

\bibitem{x11} R. H. Liu, T. Wu, G. Wu, H. Chen, X. F. Wang, Y. L. Xie, J.
J. Yin, Y. J. Yan, Q. J. Li, B. C. Shi, W. S. Chu, Z. Y. Wu, and
X. H. Chen, arXiv:0810.2694.



\end{thebibliography}
\end{document}